\newif\iflncs
\lncsfalse
\newif\ifnotes
\notesfalse

\iflncs
\documentclass[orivec,runningheads,envcountsect,envcountsame,a4paper]{llncs}
\else
\documentclass[11pt]{article}
\fi

\iflncs
\usepackage{breakcites}
\else
\usepackage{amsthm}
\usepackage{fullpage}
\fi

\usepackage{subcaption}
\usepackage{algorithm}
\usepackage[noend]{algpseudocode}
\usepackage{amsfonts}
\usepackage{amsmath}
\usepackage{bbm}
\usepackage{complexity}
\usepackage{mdframed}
\usepackage{framed}
\usepackage{float}
\usepackage{array}
\usepackage{enumitem}
\usepackage[dvipsnames]{xcolor} 
\usepackage{graphicx}
\usepackage{tikz}
\usepackage{multirow}
\usepackage{pdfpages}
\iflncs
\usepackage[colorlinks=true,linkcolor=blue,citecolor=blue]{hyperref}
\else
\usepackage[colorlinks=true,linkcolor=Maroon,citecolor=Maroon]{hyperref}
\fi
\usepackage{thm-restate}
\usepackage{orcidlink}                    
\usepackage[capitalize]{cleveref}  

\newcommand{\remove}[1]{}

\ifnotes
\newcommand{\knote}[1]{[{\footnotesize \textsf{\color{purple}{\bf Kel Zin:} { {#1}}}}]}
\definecolor{ao}{rgb}{0.0, 0.5, 0.0}
\newcommand{\pnote}[1]{[{\footnotesize \textsf{\color{cyan}{\bf Prashant:} { {#1}}}}]}
\else
\newcommand{\knote}[1]{}
\definecolor{ao}{rgb}{0.0, 0.5, 0.0}
\newcommand{\pnote}[1]{}
\fi
\definecolor{mygreen}{RGB}{0,128,0} 
\newcommand{\resolved}[1]{{\color{mygreen}{[Resolved]}}}

\iflncs
\input{lncs_issues}

\let\manualqed\qed
\spnewtheorem{claim}[theorem]{Claim}{\bfseries}{\itshape}
\else
\let\manualqed\relax
\newtheorem{theorem}{Theorem}[section]

\newtheorem{claim}[theorem]{Claim}
\newtheorem{corollary}[theorem]{Corollary}

\newtheorem{lemma}[theorem]{Lemma}

\newtheorem{definition}[theorem]{Definition}
\theoremstyle{definition}
\newtheorem{remark}[theorem]{Remark}
\fi

\iflncs
\renewcommand{\paragraph}[1]{\;\newline \noindent \textbf{#1}}
\fi

\newenvironment{proofsketch}{\begin{proof}[\textit{Proof Sketch}]}{\end{proof}}

\crefname{definition}{Definition}{Definitions}
\crefname{sub-definition}{Definition}{Definitions}
\crefname{example}{Example}{Examples}
\crefname{exercise}{Exercise}{Exercises}
\crefname{property}{Property}{Properties}
\crefname{question}{Question}{Questions}
\crefname{solution}{Solution}{Solutions}
\crefname{theorem}{Theorem}{Theorems}
\crefname{informaltheorem}{Theorem}{Theorems}
\crefname{proposition}{Proposition}{Propositions}
\crefname{problem}{Problem}{Problems}
\crefname{lemma}{Lemma}{Lemmas}
\crefname{conjecture}{Conjecture}{Conjectures}
\crefname{corollary}{Corollary}{Corollaries}
\crefname{fact}{Fact}{Facts}
\crefname{claim}{Claim}{Claims}
\crefname{remark}{Remark}{Remarks}
\crefname{note}{Note}{Notes}
\crefname{figure}{Figure}{Figure}
\crefname{case}{Case}{Cases}
\crefname{proofsketch}{Proof Sketch}{Proof Sketches}

\def\ddefloop#1{\ifx\ddefloop#1\else\ddef{#1}\expandafter\ddefloop\fi}

\def\ddef#1{\expandafter\def\csname bb#1\endcsname{\ensuremath{\mathbb{#1}}}}
\ddefloop ABCDEFGHIJKLMNOPQRSTUVWXYZ\ddefloop

\def\ddef#1{\expandafter\def\csname c#1\endcsname{\ensuremath{\mathcal{#1}}}}
\ddefloop ABCDEFGHIJKLMNOPQRSTUVWXYZ\ddefloop

\def\ddef#1{\expandafter\def\csname v#1\endcsname{\ensuremath{\overline{#1}}}}
\ddefloop ABCDEFGHIJKLMNOPQRSTUVWXYZabcdefghijklmnopqrstuvwxyz\ddefloop

\newcommand{\algfont}[1]{\mathsf{#1}}
\def\ddef#1{\expandafter\def\csname alg#1\endcsname{\ensuremath{\algfont{#1}}}}
\ddefloop ABCDEFGHIJKLMNOPQRSTUVWXYZ\ddefloop

\newcommand{\langfont}[1]{\mathnormal{#1}}
\def\ddef#1{\expandafter\def\csname lang#1\endcsname{\ensuremath{\langfont{#1}}}}
\ddefloop ABCDEFGHIJKLMNOPQRSTUVWXYZ\ddefloop

\def\ddef#1{\expandafter\def\csname v#1\endcsname{\ensuremath{\boldsymbol{\csname #1\endcsname}}}}
\ddefloop {alpha}{beta}{gamma}{delta}{epsilon}{varepsilon}{zeta}{eta}{theta}{vartheta}{iota}{kappa}{lambda}{mu}{nu}{xi}{pi}{varpi}{rho}{varrho}{sigma}{varsigma}{tau}{upsilon}{phi}{varphi}{chi}{psi}{omega}{Gamma}{Delta}{Theta}{Lambda}{Xi}{Pi}{Sigma}{varSigma}{Upsilon}{Phi}{Psi}{Omega}{ell}\ddefloop

\newcommand{\eps}{\varepsilon}
\newcommand{\Nat}{\mathbb{N}}

\newcommand{\Field}[1]{\mathbb{F}_{#1}}
\newcommand{\F}{\mathbb{F}}

\newcommand{\ra}{\rightarrow}

\newcommand{\Ot}{\tilde{O}}

\DeclareMathOperator*{\Exp} {{\mathbf{E}}}

\newcommand{\pr}[1]         {\Pr\left[ #1 \right]}
\newcommand{\prob}[2]       {\Pr_{#1}\left[ #2 \right]}
\let\exponential\exp
\renewcommand{\exp}[1]      {\Exp\left[ #1 \right]}
\newcommand{\expec}[2]      {\Exp_{#1}\left[ #2 \right]}

\renewcommand{\poly}{\mathrm{poly}}
\renewcommand{\polylog}{\mathrm{polylog}}

\newcommand{\set}[1]        {\left\{ #1 \right\}}
\newcommand{\abs}[1]        {\left| #1\right|}

\newcommand{\floor}[1]      {\left\lfloor #1 \right\rfloor}

\newcommand{\ip}[1]         {\langle #1 \rangle}

\usepackage{listofitems}
\newcommand\cycle[2][\,]{%
  \readlist\thecycle{#2}%
  (\foreachitem\i\in\thecycle{\ifnum\icnt=1\else#1\fi\i})%
}

\title{Improved Search-to-Decision Reduction for Random Local Functions}

\iflncs
\author{Kel Zin Tan\inst{1}\thanks{\orcidlink{0009-0008-1099-0892} Department of Computer Science, National University of Singapore. Email: \href{mailto:kelzin@u.nus.edu}{kelzin@u.nus.edu}} \and Prashant Nalini Vasudevan\inst{1}\thanks{\orcidlink{0000-0001-6880-795X} Department of Computer Science, National University of Singapore. Email: \href{mailto:prashvas@nus.edu.sg}{prashvas@nus.edu.sg}}}
\institute{National University of Singapore}
\date{}
\else
\author{Kel Zin Tan\thanks{\orcidlink{0009-0008-1099-0892} Department of Computer Science, National University of Singapore. Email: \href{mailto:kelzin@u.nus.edu}{kelzin@u.nus.edu}} \and Prashant Nalini Vasudevan\thanks{\orcidlink{0000-0001-6880-795X} Department of Computer Science, National University of Singapore. Email: \href{mailto:prashvas@nus.edu.sg}{prashvas@nus.edu.sg}}}
\date{\today}
\fi

\begin{document}

\pagenumbering{roman}

\maketitle

\thispagestyle{empty}

\begin{abstract}
  A random local function defined by a $d$-ary predicate $P$ is one where each output bit is computed by applying $P$ to $d$ randomly chosen bits of its input. These represent natural distributions of instances for constraint satisfaction problems. They were put forward by Goldreich~\cite{Gol11} as candidates for low-complexity one-way functions, and have subsequently been widely studied also as potential pseudo-random generators.
  
  We present a new search-to-decision reduction for random local functions defined by any predicate of constant arity. Given any efficient algorithm that can distinguish, with advantage $\eps$, the output of a random local function with $m$ outputs and $n$ inputs from random, our reduction produces an efficient algorithm that can invert such functions with $\tilde{O}(m(n/\eps)^2)$ outputs, succeeding with probability $\Omega(\eps)$. This implies that if a family of local functions is one-way, then a related family with shorter output length is a family of pseudo-random generators.
  
  Prior to our work, all such reductions that were known required the predicate to have additional sensitivity properties, whereas our reduction works for any predicate. Our results also generalise to some super-constant values of the arity $d$, and to noisy predicates.
\end{abstract}


\newpage

\iflncs
\else
\setcounter{tocdepth}{2}
\tableofcontents
\newpage
\fi

\pagenumbering{arabic}

\section{Introduction}
\label{sec:intro}

Local cryptographic primitives are those that can be computed in constant parallel time, i.e., implementable by constant-depth circuits (NC$^0$) \cite{CM01, MST03, AIK06, Gol11}. At its core, local cryptography poses the fundamental question of whether secure cryptographic functions can be designed such that each output bit depends on only a constant number of input bits. Beyond their natural efficiency, concepts from local cryptography have found many other applications, such as in secure computation \cite{ADINZ17, BCGIO17, BCM23, BCMPAR24}, learning theory \cite{DV21}, and indistinguishability obfuscation \cite{LV17, JLS21, JLS22}.

\paragraph{Random Local Functions} A remarkable natural candidate for local One-Way Functions (OWF) is the family of random local functions, introduced by Goldreich~\cite{Gol11}. Fix a $d$-ary predicate $P: \Field{2}^{d} \to \Field{2}$ for some constant $d \geq 3$. A random local function defined by $P$ is one where each output bit is set to be the result of applying $P$ to $d$ randomly chosen input bits.

To be more precise, this is a function $f_{G,P}:\Field{2}^n\ra\Field{2}^m$ such that on an input $s \in \Field{2}^n$, the $i$-th output bit of the function $f_{G,P}(s)_{i}$ is $P(s_{j_1}, \dots, s_{j_d})$, where each $j_k \in [n]$ is a randomly selected index. The collection of indices for each output bit is represented as a tuple (or an ordered hyperedge) $S_i = (j_1, \dots, j_d) \in [n]^d$. The collection of tuples is described as a hypergraph $G = (S_i)_{i \in [m]}$. Goldreich conjectured that if the predicate $P$ is appropriately chosen and $m$ is not too large or small, then such a function is one-way with high probability.\footnote{Goldreich originally studied a deterministic version of these functions where the hypergraph is fixed to be one with certain expansion properties. Nevertheless, as random graphs are expanding with high probability, his conjectures also imply the one-wayness of the functions we describe here.} 

Crucial to the hardness of inverting these functions is the choice of the predicate $P$ and the relationship between the output and input lengths $m$ and $n$. For instance, it is known that, regardless of the choice of predicate $P$, inversion can be performed efficiently for some $m = O(n^{\frac{1}{2}\floor{2d/3}}\log n)$~\cite[Corollary 3.6]{App16}. Over the past few decades, there has been an active line of work studying the complexity of inversion for various predicates and parameter settings \cite{Gol11, AHI05, BQ09, CEMT09, Its10, OW14, AL16, App16, CDMRR18, COST19, YGJL22, DMR23}. We discuss this work in more detail in \cref{sec:related}.

\paragraph{Related Problems} While conceptually simple, the problem of inverting a random local function is related to several important problems from various areas of Computer Science. Viewing the outputs of the function as constraints that the input needs to satisfy, this is an instance of a random constraint satisfaction problem (CSP) \cite{AIK08, CEMT09, RRS17, GHKM23, DV21}. 

Furthermore, suppose the predicate $P$ is a noisy XOR function, i.e. $P$ computes the parity of its inputs and flips it with a small probability. This corresponds to the sparse learning parity with noise problem (LPN) \cite{Fei02, Ale03, FGKP06, CSZ24, BBTV24}, which has many applications in public key cryptography \cite{ABW10}, machine learning \cite{BEGKMZ22}, indistinguishability obfuscation \cite{RVV24} and more \cite{CHKV24, DJ24, ABCM25, BCM25}. If we then interpret the hypergraph as describing the generator matrix of a linear code, inverting the local function corresponds to decoding its noisy codewords from random noise. Similar to how Linear Regression is usually interpreted, this may also be seen as a learning problem where the task is to learn the hidden linear function $\ip{x,\cdot}$ given many noisy evaluations.

\paragraph{Decision Problem} Interpreting the problem of inversion as a \emph{search} problem, one can naturally define the corresponding \emph{decision} problem as distinguishing the output of a random local function $(G,f_{G,P}(s))$, on a random input $s$, from $(G,b)$, where $b \in \Field{2}^m$ is uniformly random and independent of $G$. Viewing the input $s$ as the random seed, the hardness of the decision problem would imply that the family of local functions is a family of Pseudo-Random Generators (PRG).

Given their locality, such PRGs could be very efficient to compute, but have a couple of limitations. For example, with probability $1/\poly(n)$, the graph $G$ has a pair of repeated hyperedges, in which case the corresponding output bits of $f_{G,P}$ can be used to distinguish it from random. The best bound on an adversary's distinguishing advantage we can hope for against a random local function is thus some $1/\poly(n)$; we refer to such PRGs as ``weak'' PRGs. Second, known algorithms place a limit of $\Ot(n^{\floor{2d/3}/2})$ on the stretch of this PRG if a $d$-ary predicate is used~\cite{App16}, and even stronger limits are known for certain classes of predicates~\cite{BQ09,AL16}.

Regardless, it is still possible to construct a local PRG with negligible distinguishing advantage (a ``strong'' PRG) from a random local function by processing the output further. Applebaum~\cite{App12} showed how to get a strong local PRG with linear stretch by applying a local extractor to the output of an unpredictable random local function. Later, Applebaum and Kachlon~\cite{AK19} effectively resolved these issues by showing that any weak local PRG that has non-trivial polynomial stretch can be transformed into a strong local PRG with any desired polynomial stretch, at the cost of some degradation of the locality. The key idea is to apply an additional layer of a local randomness extractor to the output of the weak PRG.

\begin{theorem}[{\cite[Theorem 2.12]{AK19}} weak-to-strong compiler]
    \label{thm:ak19}
    For every constant $d \in N$, $a > 0$ and $c, c^\prime > 1$, there exists a constant $d^\prime$ for which the following holds. Any $d$-local PRGs that stretch $n$ bits into $n^c$ bits with at most $\eps = 1/n^a$ distinguishing advantage can be converted into $d^\prime$-local PRGs that stretch $n$ bits to $n^{c^\prime}$ bits with at most a negligible distinguishing advantage.
\end{theorem}

\paragraph{Search-to-Decision Reductions} Following this discussion, if we can reduce the above search problem to the decision problem, we can construct local PRGs from the one-wayness of random local functions. Further, given the fundamental nature of the family of local functions and the aforementioned connections to CSPs, decoding, etc., such reductions would be of broad interest even outside the context of local cryptography.

Such a reduction was presented by Applebaum~\cite{App12} for predicates $P$ that satisfy a sensitivity requirement. We say that a predicate $P$ is sensitive if there exists a variable such that flipping this variable always results in the flipping of the output. 

\begin{theorem}[{\cite[Theorem 1.4]{App12}}]
    \label{thm:weakprg}
    Fix any sensitive $d$-ary predicate $P$. Suppose there exists an efficient algorithm with distinguishing advantage $\eps$ for the decision problem for random local functions defined by $P$ with $m$ outputs and $n$ inputs. Then there is an efficient algorithm with success probability $\Omega(\eps)$ for the search problem for random local functions defined by $P$ with $O(m^3/\eps^2)$ outputs and $n$ inputs. 
\end{theorem}

They extend the above result to accommodate noisy predicates $P$, incurring an additional $\log n$ factor in the output length required by the search algorithm~\cite[Theorem 4.5]{App12}. The proof uses a distinguisher to derive a ``next-bit predictor'' that can predict an output bit given all previous output bits. It is then subsequently used to solve the search problem.

They also show that if the next-bit predictor gives a constant advantage, then the function can be inverted with constant probability; this holds without any blowup in output length and without requiring the predicate to be sensitive, but falls short of leading to pseudo-randomness or a reduction to the decision problem.

There have since been a few notable improvements to the above search-to-decision reduction in the special case of $P$ being the noisy XOR predicate. Assuming the existence of a decision algorithm for $m$ outputs and $n$ inputs with $\eps$ distinguishing advantage for the noisy $d$-ary XOR predicate $P$ with noise parameter $\eta$,
\begin{itemize}[itemsep=0pt]
    \item \cite{BSV19} showed that there exists an algorithm for the search problem with $O(m(m/\eps)^{2/d})$ outputs, maintaining the same locality constant $d$, but succeeding only with small probability $\exponential (-\tilde{O}(m/\eps)^{6/d})$. Their approach relies on a new approximate local list-decoding algorithm for the $d$-XOR code at large distances.
    \item \cite{BRT25} showed that there exists an algorithm for the search problem with $O(nm + n/\eta^2\eps^{2(d-2)})$ outputs and $n$ inputs for the noisy $(d+1)$-ary XOR predicate. They further generalise their result to any sensitive predicate, though again the predicates in the decision and search problems are slightly different. The main approach is a transformation that removes a sensitive variable from the predicate, creating a distributional shift. This shift then allows the construction of a predictor that solves the search problem.
\end{itemize}

\subsection{Our Contribution}

In this work, we present a new search-to-decision reduction for random local functions, captured by the following theorem.

\begin{theorem}[Informal, see \cref{thm:general}]
  \label{infthm:main}
  Fix any $d$-ary predicate $P$. Suppose there exists an efficient algorithm with distinguishing advantage $\eps$ for the decision problem for random local functions defined by $P$ with $m$ outputs and $n$ inputs. Then there is an efficient algorithm with success probability $\Omega(\eps)$ for the search problem for random local functions defined by $P$ with $O(m (n/\eps)^2 \log^3(n/\eps))$ outputs and $n$ inputs.
\end{theorem}

\begin{remark}
    If the predicate $P$ is biased, the distinguishing task with a random binary string is trivial. In this case, we define the decision problem to instead distinguish from a random binary string with the same bias as $P$. That is, if $\Pr[P(x) = 1] = \eta$, then the distinguishing task is with the distribution $(G, b)$ where $b$ is sampled from $Bern(\eta)^m$.
\end{remark}

The main improvement of our work over all existing search-to-decision reductions is the removal of the requirement of the predicate being sensitive. This generalisation opens up the possibility of constructing local PRGs (including strong ones, following \cref{thm:ak19}) using the hardness of inverting a wider range of local functions. Lack of sensitivity is potentially beneficial from a cryptographic perspective, as it means one less form of structure, which may make certain classes of attacks harder to mount.

While some of the steps in our reduction are similar to those in prior work, the core of our approach is substantially different. For instance, in contrast to \cite{App12,BSV19}, we directly obtain our search algorithm from the decision algorithm, without going through an intermediate predictor for output bits. And in contrast to \cite{BRT25}, the predicate is preserved in the course of our reduction.

Our analysis depends only minimally on the predicate itself, relying instead on the mixing properties of certain transformations we perform on the hypergraph. We hope that these techniques will be useful in similar reductions beyond the context of local functions as well. We note that our reduction does not achieve the same level of sample complexity as those for sensitive predicates (approximately losing a factor of $n$), but we introduce techniques that we hope will allow future work to close the remaining gap between our loss factor and the tighter bounds known for sensitive predicates.

\paragraph{Optimal Reduction} For sensitive predicates, there is currently no proven generic sample complexity gap between the search problem and the decision problem. For a general predicate, however, there is a trivial gap of $\sqrt{n}$. Consider a random local function with $m$ outputs and $n$ inputs, with the predicate $P$ being the majority of 3 variables. One can solve the distinguishing task by finding a pair of outputs that depend on the same input. This is because for a majority predicate, sharing an input implies correlated output. With around $m = \sqrt{n}$ samples, such a pair appears with constant probability. However, it is information theoretically impossible to solve the search with just $\sqrt{n}$ outputs. Therefore, the best possible search to decision reduction reduces search with $\tilde{O}_{\eps}(m\sqrt{n})$ samples to decision with $m$ samples. This shows that our reduction is $\tilde{O}_{\eps}(n^{1.5})$ factor far from optimal.

\paragraph{Generalisation} In our hypergraph modelling of local functions, we allow a hyperedge to contain repeated vertices. The alternative model that requires all vertices in a hyperedge to be distinct is also commonly used. We extend our reduction to work for this model as well (see \cref{sec:distinct-hyperedge}). We also extend it to handle larger locality $d = \polylog(n)$, under some additional restrictions on $m$ and the advantage $\eps$ of the distinguisher (see \cref{sec:non-constant-sparsity}). Finally, we can also handle noisy predicates where independent Bernoulli noise is added to the output (see \cref{sec:noisy-predicate}).

\subsection{Overview of Techniques} 
\label{sec:overview}

In this subsection, we provide an overview of the techniques used to perform our search-to-decision reduction, starting with the basic definitions. For the rest of the section, fix some $d$-ary predicate $P$ with $\eta = \prob{x}{P(x)=1}$.

\paragraph{Notation} Given a binary string $x$, we use $x_i$ to denote the $i$-th bit in the binary string. Denote by $G_{n,m,d}$ the set of all hypergraphs with $n$ vertices and $m$ ordered hyperedges, with $d$ vertices in each hyperedge. For a distribution $D$, we denote $x \gets D$ to indicate that $x$ is sampled from $D$; if $D$ is a set, this denotes sampling uniformly from the set. For any $\eta\in [0,1]$, we denote by $Bern(\eta)$ the Bernoulli distribution with parameter $\eta$. We say that a distinguishing algorithm $\algD$ has advantage $\eps$ in the decision problem with predicate $P$ if:
\[
    \prob{\substack{G\gets G_{n,m,d} \\ s\gets \Field{2}^n}}{\algD(G, f_{G,P}(s)) = 1} - \prob{\substack{G\gets G_{n,m,d} \\ b\gets Bern(\eta)^m}}{\algD(G,b) = 1} \geq \eps
\]

\paragraph{Reduction Idea} The high-level idea of our reduction is as follows. We start with a distinguisher $\algD$ with advantage $\eps$ as above. We use it to construct a set of predictor algorithms $\algS_2,\dots,\algS_n$, where each $\algS_i$ has a small advantage $\Omega(\eps/t)$ in predicting the value of $s_1\oplus s_i$, for some $t = O(n\log(n/\eps))$. Further, we show that for an $\Omega(\eps)$ fraction of secrets $s$, all of the $\algS_i$'s simultaneously express this advantage. Whenever such a secret happens to be selected, we can then amplify all these advantages using $O((t/\eps)^2\log{n})$ independent samples and learn all the parities $s_1\oplus s_i$ with high confidence, thus recovering $s$ itself. Altogether, the search algorithm needs $O((n/\eps)^2\log^3(n/\eps))$ independent instances, each with $m$ outputs, and succeeds with probability $\Omega(\eps)$, as stated in \cref{infthm:main}.

\paragraph{Predictor} We will first provide more details on the construction of the predictors. Suppose we have a decision algorithm $\algD$ that has advantage $\eps$. Our goal will be to construct a set of algorithms $\algS_i$, such that for a large enough fraction of secrets $s \in \Field{2}^n$, there is a pair of numbers $eq_s, neq_s \in [0,1]$ where $eq_s > neq_s$, such that the following holds for all $i\in[2,n]$:

\begin{itemize}[itemsep=0pt]
    \item If $s_1 = s_i$, then $\Pr[\algS_i(G, f_{G,P}(s)) = 1] = eq_s$
    \item If $s_1 \neq s_i$, then $\Pr[\algS_i(G, f_{G,P}(s)) = 1] \leq neq_s$
\end{itemize}
As long as the gap between $eq_s$ and $neq_s$ is not negligible, we can tell the relationship between $s_1$ and $s_i$ by running the algorithm $\algS_i$ polynomially many times with independent problem instances with the same secret $s$. 

Each $\algS_i$, on input some $(G,y)$, functions by applying a randomised transformation $T$ (described later) to the hypergraph $G$. This transformation is designed such that if $s_1 = s_i$, $(T(G), f_{G,P}(s))$ looks more like a random local function instance $(G, f_{G,P}(s))$, and if $s_1 \neq s_i$, it looks more like $(G, b)$ where $b \gets Bern(\eta)^m$. Next, the transformed problem instance $(T(G),y)$ is fed to the decision algorithm $\algD$, and $\algS_i$ outputs whatever $\algD$ does. We will show using a hybrid argument that, for any secret $s$ for which $\algD$ still has $\Omega(\eps)$ advantage in distinguishing between $(G,f_{G,P}(s))$ and $(G,b)$, the gap between $eq_s$ and $neq_s$ is at least $\Omega(\eps/t)$, where $t = O(n\log(nm/\eps)) = O(n\log(n/\eps))$ (since $m = poly(n)$).

\paragraph{Transformation} The key to our improvement comes from the transformation. Here, we will show how we achieve the gap of $\eps/t$ between $eq_s$ and $neq_s$ where $t = O(n\log(nm/\eps))$.

The transformation $T_{a,b} : G_{n,m,d} \to G_{n,m,d}$ is a randomized function parameterised by two values $a, b \in[n]$ that takes in a hypergraph $G = (S_i)_{i \in [m]}$ with $S_i = (j_1, \dots, j_d)$ and returns another hypergraph $G^\prime = (S^\prime_i)_{i \in [m]}$ with $S_i^\prime = (j^\prime_1, \dots, j^\prime_d)$, where the distribution of each of the $j_k^\prime$ is given as follows
\[
            \Pr[j_k^\prime = j_k] = 1 \quad\text{ if } j_k \notin \{a, b\}\\
\]
\[
    \Pr[j_k^\prime = a] = \dfrac{1}{2},\ \Pr[j_k^\prime = b] = \dfrac{1}{2}  \quad\text{ if } j_k \in \{a, b\}
\]
Essentially, if a vertex in the hyperedge is not $a$ or $b$, then it will remain. Otherwise, the vertex will change to either $a$ or $b$ with probability half. It is easy to see that the uniform distribution over $G_{n,m,d}$ is stable under this process, i.e the uniform distribution of $G_{n,m,d}$ is the same as the distribution of uniformly sampling from $G_{n,m,d}$ and then applying the transformation. 

\medskip
Observe that if $s_a = s_b$, then the output of the random local function remains the same after the transformation: $f_{G,P}(s) = f_{T_{a,b}(G),P}(s)$. The transformation does nothing to the distribution of $(G, f_{G,P}(s))$ when $G$ is uniformly distributed. Using $\approx$ to denote similarity between distributions, in this case, we have:
\[ (T(G), f_{G,P}(s)) = (T(G), f_{T(G),P}(s)) \approx (G, f_{G,P}(s))  \]
On the other hand, when $s_a \neq s_b$, then in general $f_{G,P}(s)$ is not the same as $f_{T_{a,b}(G),P}(s)$, and the relationship between $T(G)$ and $f_{G,P}(S)$ becomes less coupled than between $G$ and $f_{G,P}(s)$. In fact, we can show that after we apply about $t = O(n\log(nm/\eps))$ such transformations, the transformed hypergraph will become independent of $f_{G,P}(s)$.

Precisely, we show that for any hypergraph $G$, the distribution of $T_{a_t,b_t}\circ \ldots \circ T_{a_1, b_1}(G)$ is close to uniformly sampling from $G_{n,m,d}$. One can think of the transformation as a Markov process, and $O(n \log (nm/\eps))$ is the mixing time of this process.\footnote{The proof of this claim is inspired by the convergence of randomised gossip algorithms \cite{BGPS06}.} This implies that with an independently sampled $G^\prime \gets G_{n,m,d}$,
\[
    (T_{a_t,b_t}\circ \ldots \circ T_{a_1, b_1}(G), f_{G,P}(s)) \approx (G^\prime, f_{G,P}(s))
\]
To see why this might be true, note that after about $t = \Omega(n \log (nm/\eps))$ such randomly chosen transformations, every vertex in every hyperedge of $G$ would have been touched by at least a few of the transformations. As the random assignment in $T_{a,b}$ happens independently for each $a$ or $b$ that appears as a vertex, this process quickly randomises the whole hypergraph.

\medskip
We can, in fact, show an even stronger result: after applying $t = O(n \log(nm/\eps))$ transformations, the transformed instance $(T(G),f_{G,P}(s))$ resembles an instance $(G, b)$, where $b$ is sampled from $Bern(\eta)^m$. The main task here is to prove $f_{G,P}(s) \approx Bern(\eta)^m$ when $G$ is sampled at random. Notice that as $G$ itself is not revealed (being completely hidden by $T(G)$), each bit of $f_{G,P}(s)$ is simply an independent output of $P$ when its inputs are chosen to be random bits from $s$. The bias of this output depends heavily on the Hamming weight of $s$. 

If $s$ is extremely biased towards $1$ or $0$, one might be able to distinguish between these $f_{G,P}(s)$ and $b$ with just a small output size $m$. Fortunately, for a randomly sampled $s$, the Hamming weight is usually fairly balanced, around $n/2 \pm O(\sqrt{n})$, and the bias of $P$ on this input distribution remains close to $\eta$. We then claim that, conditioned on $s$ being fairly balanced, one will need a large number of output bits to distinguish between $f_{G,P}(s)$ and $b$. And when the number of outputs reaches a point where this distinguishing is possible, both the search and decision problems turn out to be easy due to known algorithms from \cite{App16}. 

Therefore, when $m$ is not so large that the search problem is already easy, applying $t$ transformations to the hypergraph $G$ makes the joint distribution of the transformed hypergraph and $f_{G,P}(s)$ look like $(G, b)$ where $b \gets Bern(\eta)^m$. So when we apply the transformation $T_{a, b}$ on $s_a \neq s_b$, the distribution is indeed a step closer to the $(G, b)$ distribution. The algorithm $\algS_i$, on input $(G,y)$, applies a random number $r\gets[0,t-1]$ of transformations $T_{a,b}$ to $G$ with random $a$, $b$, and then finally applies the transformation $T_{1,i}$ once. Then, by a straightforward hybrid argument, we obtain the properties of $\algS_i$ listed earlier, including that the gap between $eq_s$ and $neq_s$ is at least $\Omega(\eps/t)$.

There are a few issues that arise from the possibility of the distinguisher's behaviour being correlated with the secret that we have glossed over here. These are readily dealt with by simple transformations like permuting the secret first, and are accounted for in the actual proof.

\subsection{Related Work}
\label{sec:related}

\paragraph{Security of Random Local Functions} Goldreich \cite{Gol11} originally proposed using a randomly chosen predicate, while cautioning against linear, degenerate, or otherwise structured predicates that lead to easily solvable equation systems. It is known that myopic and drunken backtracking algorithms (algorithms that only look at the instance locally each time they make a decision; this includes many powerful SAT solvers) do not perform well on predicates with some linear component $P(x_1, \dots, x_d) = x_1 + \dots + x_k + Q(x_{k+1}, \dots, x_{d})$ for an arbitrary $Q$ with $k > 3d/4$ for some large enough $d$ on $n$ variables and $n$ outputs \cite{AHI05, CEMT09, Its10}. For the XOR-AND$_{3,2}$ predicate $P(x_1, \dots x_5) = x_1 + x_2 + x_3 + x_4x_5$, \cite{OW14} show pseudorandomness up to output length $n^{1.499}$ against $\Field{2}$-linear tests and a wide class of semi-definite programming algorithms. The work of \cite{ABR12} showed a dichotomy -- for output length $n^{1+\delta}$, every choice of the predicate results in the random local function that either is secure against linear tests with high probability, or is insecure with high probability.

On the negative side, \cite{BQ09} presented attacks against predicates that exhibit strong correlation with one or two input variables. \cite{AL16} showed necessary conditions on the predicate $P$ to prevent certain families of algebraic attacks. For a random local function with predicate $P$, and output length $m=n^s$, they show that it is necessary to be $\Omega(s)$-resilient ($s$-resilient means that $P$ is uncorrelated with any $s$-subset of its input) and have an algebraic degree of $\Omega(s)$ even after fixing $\Omega(s)$ of its inputs. In particular, they show that for all $\ell, k$, the XOR-AND$_{\ell,k}$ predicate suggested by $\cite{OW14}$ is not pseudorandom with $n^{2.01}$ outputs. They then suggest an alternative candidate, XOR-MAJ$_{\ell,k}$. \cite[Corollary 3.6]{App16} showed that regardless of the choice of predicate $P$, observing more than $\Omega(n^{\frac{1}{2}\floor{2d/3}}\log n)$ outputs gives an efficient algorithm to recover the input.

\cite{CDMRR18} looked at the concrete security of Goldreich's function and developed a sub-exponential time attack and analysed the efficiency of algebraic attacks, such as Gröbner basis methods. \cite{YGJL22} built on their work, improving the time complexity. \cite{COST19} showed that the assumption of the hypergraph being a good expander is not sufficient; the neighbour function of the graph must also have high circuit complexity. 

More recently, \cite{DMR23} investigated the problem of designing predicates that simultaneously achieve high resilience and high algebraic immunity. A predicate with high algebraic immunity guarantees a high algebraic degree after fixing some inputs, though the converse does not necessarily hold \cite{AL16}. \cite{DMR23} conjectured the existence of such optimal predicates and demonstrated that the commonly used XOR-MAJ predicate does not satisfy this optimality criterion. Furthermore, through experiments, they identified optimal predicates for localities up to 12.

On attacking local PRGs, it is known that there exists a subexponential time algorithm for distinguishing superlinear stretch local PRGs with noticeable probability \cite{AIK06, Zic17, Una23}. A related variant of local PRG is also studied by \cite{ABCM25}, where the input string is not uniformly sampled. Assuming the hardness of sparse LPN, they showed that superlinear stretch with negligible distinguishing advantage is possible.

\paragraph{Local Cryptography} Cryan and Miltersen \cite{CM01} initiated the study of local PRGs, where they showed the impossibility of a PRG in NC$^0_3$ (circuit with depth of $3$) with superlinear stretch. Mossel et al. \cite{MST03} then extended the impossibility to NC$^0_4$. 

On the positive side, Applebaum et al.~\cite{AIK06} proved the existence of an OWF and a sublinear stretch PRG in NC$^0$, assuming the existence of an OWF and a PRG in NC$^1$ (logarithmic depth circuits), which follows from well-established cryptographic assumptions such as the hardness of lattice problems. The same authors \cite{AIK08} further showed the existence of a linear-stretch local PRG assuming the hardness of the average case MAX-3LIN problem \cite{Ale03}, which is related to the sparse LPN problem. 

Goldreich \cite{Gol11} proposed the local OWF candidate discussed earlier in this section, based on random local functions. It is known that under this assumption, one can obtain a polynomial stretch local PRG from the work of \cite{App12, AK19} and locally computable universal one-way hash functions with linear shrinkage \cite{AM13}. Other related work includes the hardness amplification results for local OWFs by \cite{BR11}.

\subsection*{Organization} In \cref{sec:defs}, we describe preliminaries and formally define the terms that we will use throughout the paper. \cref{sec:general} will be on the proof of our main search-to-decision reduction theorem. \cref{sec:generalization} will discuss the generalisation of our theorem to other interesting families of problems. Finally, in \cref{sec:deferred}, we include deferred proofs and additional details from the other sections.



\section{Preliminaries}
\label{sec:defs}

\paragraph{Notation.} Given a binary string $x$, we use $x_i$ to denote the $i$-th bit in the binary string. Given two distributions $D_1, D_2$, If the two distributions are equivalent, we write $D_1 \approx D_2$. If $x \gets D_1$, this means $x$ is sampled from $D_1$. Here, $D_1$ can also be a set, in which case this denotes sampling uniformly from the set. We denote a finite field of size $p$ as $\Field{p}$ where $p$ is a prime. For any $\eta\in [0,1]$, we denote by $Bern(\eta)$ the Bernoulli distribution with parameter $\eta$.

\begin{definition}[Statistical Distance]
  \label{def:statistical-distance}
    Consider two distributions $D_1, D_2$ defined over space $Q$, the statistical distance (Total Variation Distance) is defined as
    \[
        \Delta(D_1, D_2) = \dfrac{1}{2} \sum_{x \in Q}\left|\Pr[D_1 = x] - \Pr[D_2 = x]\right|
    \]
\end{definition}

\begin{definition}[Hamming weight]
    Denote the Hamming weight of a binary string $x$ as $wt(x)$.
\end{definition}

\begin{definition}[$d$-ary]
      For $d \in \Nat$, a \emph{$d$-ary} predicate $P$ is a function $P: \Field{2}^d \to \Field{2}$ (that is, it takes $d$ elements as input).
\end{definition}

\begin{definition}[Bias]
    Suppose a \emph{$d$-ary} predicate P, the \emph{bias} of $P$ is defined as $\expec{x \gets \Field{2}^d}{P(x)} = \prob{x \gets \Field{2}^d}{P(x) = 1}$. We commonly denote the bias as $\eta \in [0, 1]$
\end{definition}

\begin{definition}[Bounded Bias]
    \label{def:bounded_bias}
    Suppose a \emph{$d$-ary} predicate P with bias $\eta$, we say that the bias is bounded if there exist two constants $c_1, c_2$ such that $0 < c_1 \leq c_2 < 1$ and $\eta \in [c_1, c_2]$
\end{definition}

\begin{definition}[$c$-correlated, {\cite[Section 3.2]{App16}}] We say that a non-constant predicate $P$ is \emph{$c$-correlated} if $c$ is the minimal positive integer such that it is correlated with the parity of a cardinality-$c$ subset of its inputs. More formally,
\[
    \Pr[P(x) = \sum_{i \in T} x_i] \neq \dfrac{1}{2}
\]
for some subset $T$ with cardinality $c$.
\end{definition}

\begin{definition}[Hypergraph]
    For $n,m,d \in \Nat$, an \emph{$(n,m,d)$-hypergraph} is a hypergraph on $n$ vertices with $m$ hyperedges, where each hyperedge $S_i$ is an ordered tuple $S_i = (j_1, \dots, j_d) \in [n]^{d}$. We commonly write a hypergraph as $G = (S_i)_{i \in [m]}$
\end{definition}

Note that this definition of hypergraph allows repeated values in the hyperedges. We denote the set of all $(n,m,d)$-hypergraphs as $G_{n,m,d}$, and also use this symbol to denote the uniform distribution over this set.

\begin{definition}[$d$-local function]
    Consider a $d$-ary predicate $P$ and a $(n,m,d)$-hypergraph $G$, a \emph{$d$-local function} is a function $f_{G,P} : \Field{2}^{n} \to \Field{2}^m$ such that given input $x$, the $i$-th output is defined as
    \[
        f_{G,P}(x)_i = P(x_{j_1}, \dots, x_{j_d})
    \]
    where $S_i = (j_1, \dots, j_d)$ is the $i$-{\text{th}} hyperedge of $G$.
\end{definition}

Informally, the $i$-th hyperedge decides which input indices are applied to the predicate to compute the $i$-th output. 

\begin{definition}[Decision Problem]
  \label{def:decision-klin}
  Consider a $d$-ary predicate $P$ with bias $\eta$ and the uniform hypergraph distribution $G_{n,m,d}$. An algorithm $\algD$ is said to have advantage $\eps \in[0,1]$ in solving the \emph{Decision Problem for $(P,n,m)$}, if the following holds:
  \begin{align*}
    \abs{ \prob{\substack{G\gets G_{n,m,d} \\ s\gets \Field{2}^n}}{\algD(G, f_{G,P}(s)) = 1} - \prob{\substack{G\gets G_{n,m,d} \\ b\gets Bern(\eta)^m}}{\algD(G,b) = 1}  } \geq \eps
  \end{align*}
\end{definition}

In other words, the algorithm can distinguish the output of the $d$-local function from a random binary string of length $m$ that has the same bias as the predicate $P$. The corresponding hypergraph of the local function is randomly sampled from $G_{n,m,d}$, and the input is randomly sampled from $\Field{2}^n$.

\begin{remark}
    Suppose a $d$-predicate $P$ with bias $\eta$, $G \gets G_{n,m,d}, s \gets \Field{2}^n, b \gets Bern(\eta)^m$. The distribution $(G, f_{G,P}(s))$ is called the \emph{planted distribution} while $(G,b)$ is called the \emph{null distribution}.
\end{remark}

\begin{definition}[Search Problem]
  \label{def:search-klin}
    Consider a $d$-ary predicate $P$ and the uniform hypergraph distribution $G_{n,m,d}$. An algorithm $\algS$ is said to have success probability $\eps \in[0,1]$ in solving the \emph{Search Problem for $(P, n, m)$} if the following holds:
  \begin{align*}
    \prob{\substack{G\gets G_{n,m,d} \\ s\gets \Field{2}^n}}{\algS(G, f_{G,P}(s)) = s^\prime, \text{ such that } f_{G,P}(s) = f_{G,P}(s^\prime)} \geq \eps
  \end{align*}
\end{definition}

\begin{remark}
    To simplify notation, we assume that the description of the predicate used $P$ is public knowledge. All algorithms have access to the description of $P$. Also, all predicates $P$ are assumed to be non-constant, as the search problem on constant predicates is trivial.
\end{remark}

\begin{lemma}[Sample threshold, {\cite[Theorem 3.5]{App16}}]
\label{thm:threshold}
Given a $c$-correlated $d$-ary predicate $P$, where $d = O(1)$, there exists a polynomial-time algorithm for both the Search and Decision Problems for $(P,n,m)$ for some $m  = O(n^{c/2} + n\log n)$ with success probability and advantage, respectively, that is $1 - o(1)$. 
\end{lemma}

\begin{proofsketch}
    Suppose $T$ is a minimal subset of size $c$ that is correlated to the predicate. Since $d = O(1)$, the correlation with the parity is a constant. Without loss of generality,
    \[ \Pr[P(x) = \sum_{i \in T} x_i] - \dfrac{1}{2}\geq 2^{-d} = \Omega(1) \] 
    Each output of the $d$-local function can then be written as parity sum + noise, 
    \[ f_{G,P}(x)_j = \sum_{i \in T}x_i + e\]
    where $e$ is a Bernoulli random variable. A constant correlation also implies that the Bernoulli parameter is constant. An algorithm to solve the search problem is constructed by reducing the problem to a $c$-sparse noisy linear system. Having $m = \Omega(n^{c/2})$ gives roughly $\Omega(n)$ pairs of equations where $c-1$ variables are the same, while the rest of the variables are disjoint. Each pair then has its equations combined to obtain 2-sparsity noisy linear equations with constant error rate. Since there exists an efficient algorithm to solve a 2-sparsity noisy linear system with a constant error rate, we are done. 
\end{proofsketch}

\section{Search-to-Decision Reduction}
\label{sec:general}

\begin{theorem}
  \label{thm:general}
  Consider a $d$-ary predicate $P$, with $d = O(1)$. Suppose there is a polynomial-time algorithm that has advantage $\eps \in [0,1]$ in solving the Decision Problem for $(P,n,m)$ for some $m = poly(n)$. Then there is a polynomial-time algorithm for the Search Problem for $(P,n,\ell m)$ that has success probability $\Omega(\eps)$, for some $\ell = \Theta((n/\eps)^2\log^3 (n/\eps))$.
\end{theorem}
    The idea of the proof is that we first show there exist $t = \Omega(n\log (n/\eps))$ hybrids $H_0, \dots, H_t$ where $H_0$ hybrid has the same distribution as the planted distribution and $H_t$ distribution has a small statistical distance from the null distribution. Then we construct a predictor on the value of $s_i$ where $i \in [2, n]$ by applying a transformation to the $H_i$ hybrid such that when $s_1 = s_i$, the transformed distribution will be $H_i$ hybrid, and $H_{i+1}$ otherwise. Using the hybrid argument, this will give a $\eps/t$ advantage in predicting the value of $s_1\oplus s_i$. This prediction can then be amplified with $\Theta((t/\eps)^2 \log (n/\eps))$ repetition and guessing the value of $s_1$ to recover the secret $s$.

    Suppose that $P$ is $c$-correlated, we can then assume that $m = o(n^{c/2 - 2})$ as if not then both Search and Decision for distribution $G_{n, k \times m, d}$ will become easy due to \cref{thm:threshold}, making our theorem statement trivially true. The same reasoning implies that we can assume $\eps = w(n^{-c/4})$. This assumption will be used in \cref{claim:random_output}

    The rest of the section will cover the proof of this theorem. It is split into 3 parts: the definition of the hybrids and proofs of their properties, the construction of the predictor algorithm, and the amplification of its advantage. Throughout the proof, fix any $d$-ary predicate $P$ for any constant $d$.
    
    \subsection{Hybrids}

    In this subsection, we will define the hybrids $H_i$ and prove that the $H_0$ hybrid has the same distribution as the planted distribution and the $H_t$ distribution has a small statistical distance from the null distribution. First, we start by explaining the hybrid argument.

    \begin{claim}[Hybrid Argument] 
    \label{claim:hybrid_argument}
    Suppose there are $t$ distributions $H_0, \dots, H_t$ such that an algorithm $\algD$ has advantage $\eps$ in distinguishing $H_0$ and $H_t$. Then the algorithm can distinguish $H_i$ and $H_{i+1}$ with advantage $\eps/t$ on a randomly chosen $i$. Formally,
    \[
        \abs{\Pr_{i \gets [0, t-1]}[D(H_i) = 1] -  \Pr_{i \gets [0, t-1]}[D(H_{i+1}) = 1]} \geq \eps/t
    \]
    \end{claim}

    \begin{definition}[Permuted Hypergraph] 
        For a permutation $\pi$ on $[n]$ and a hypergraph $G = (S_i)_{i\in[m]}$, we define the permuted hypergraph $\pi(G) = (S^\prime_i)_{i\in[m]}$ as having the values in the hyperedges permuted with $\pi$. i.e. Suppose $S_i = (j_1, \dots, j_d)$, then
        \[
            S^\prime_i = (\pi(j_1), \dots, \pi(j_d))
        \]
    \end{definition}
    Next, we define a randomised transformation on a hypergraph $G = (S_i)_{i\in[m]}$ parameterised by two values $a,b \in [n]$, such that the member value in each hyperedge $S_i$ will remain the same if it is not $a$ and not $b$. Otherwise, it will switch to $a$ or $b$ with probability half. 
    \begin{definition}[Transformation]
    \label{transformation}
      Define a \emph{randomized transformation $T_{a,b}: G_{n, m, d} \to G_{n,m,d}$} as follows: On inputting a hypergraph $G$, each of the hyperedges $S_i = (j_1, \dots, j_d)$ is transformed to $S_i^\prime = (j_1^\prime, \dots, j_d^\prime)$ independently where
      \[
            \Pr[j_k^\prime = j_k] = 1 \quad\text{ if } j_k \notin \{a, b\}\\
      \]
      \[
            \Pr[j_k^\prime = a] = \dfrac{1}{2},\ \Pr[j_k^\prime = b] = \dfrac{1}{2}  \quad\text{ if } j_k \in \{a, b\}
      \]      
    \end{definition}
    \noindent The hybrid is then defined by repeatedly applying this transformation to the permuted input hypergraph. 
    \begin{definition}[Hybrid]
        \label{def:hybrid}
        Given a secret $s$, define the \emph{$i$-th hybrid distribution $H_i^s$} as applying $i$ many randomly selected transformations to the permuted randomly sampled hypergraph of a $d$-local function.
        \[
            H_i^s = (T_{a_i,b_i} \circ \ldots \circ T_{a_1,b_1}(\pi(G)), f_{G,P}(s))
        \]
        where $G \gets G_{n,m,d}$, $\pi$ is a randomly sampled permutation on $[n]$, and for each $j\in [i]$, $a_j, b_j \gets [n]$. 
    \end{definition}

    Now we show that the terminal hybrids $H_0$ and $H_t$ are similarly distributed to the planted and the null distributions, respectively.  

    \begin{definition}[$\eps$-fairly balanced secret]
        \label{def:fairly_balanced}
        For $\eps \in [0,1]$, we say that a secret $s \in \Field{2}^n$ is \emph{$\eps$-fairly balanced} if the Hamming weight $wt(s) \in [n/2 - w, n/2 + w]$ where $w = 2\sqrt{n}\log (1/\eps)$.
    \end{definition}

    \begin{remark}
        \label{rem:fairly_balance_prob}
        By Chernoff bound, the probability that a randomly sampled secret $s \gets \Field{2}^n$ is not $\eps$-fairly balanced is at most $o(\eps)$.
    \end{remark}
  \begin{lemma}[Terminal Hybrid]
    \label{lemma:terminal_hybrid}
        Suppose $d = O(1)$, $m = o(n^{c/2 - 2})$ and $\eps = \omega(n^{-c/4})$, consider any $c$-correlated $d$-ary predicate $P$ with bounded bias $\eta$, and any $\eps$-fairly balanced fixed $s \in \F_2^n$. With the sampling $s'\gets \F_2^n$ conditioned on $wt(s) = wt(s^\prime)$ and $G\gets G_{n,m,d}$, we have the following: 
    \begin{enumerate}[itemsep=0pt]
      \item The distribution of $H_0^s$ is identical to $(G, f_{G, P}(s^\prime))$.
      \item The distribution of $H_t^s$ has statistical distance at most $\eps/4$ from $(G, b)$, where $b \gets Bern(\eta)^m$, for some $t = O(n\log (nm/\eps)) = O(n\log (n/\eps))$.
    \end{enumerate}
  \end{lemma}
  \begin{proof}
      For $H_0^s$, the only changes made are that a randomly sampled permutation is applied to the hypergraph. However, since the hypergraph is randomly sampled and $s^\prime$ is sampled such that it has the same Hamming weight as $s$, it is identically distributed to $(G, f_{G, P}(s^\prime))$. 
      \[
        H_0^s = (\pi(G), f_{G,P}(s)) = (\pi(G), f_{\pi(G),P}(\pi(s))) \approx (G, f_{G,P}(\pi(s))) \approx (G, f_{G,P}(s^\prime))
      \]
      For $H_t^s$, we claim that no matter the initial starting hypergraph $G$, after applying the transformations to $G$, it will become distributed like a randomly sampled hypergraph. Formally,
      \begin{restatable}[Random Graph]{claim}{claimrandomgraph}
          \label{claim:random_graph}
          Given any $G \in G_{n,m,d}$, if $a_j, b_j \gets [n]$, $m = poly(n)$ and for some $t = O(n\log (nm/\eps))$, then the statistical distance between $T_{a_t,b_t}\circ\ldots\circ T_{a_1,b_1} \circ G$ and the uniform distribution over $G_{n,m,d}$ is at most $\eps/8$. 
      \end{restatable}
      \begin{proofsketch}
        Intuitively, each hyperedge will get more randomised when a random transformation is applied. Once we apply $t = \Omega(n \log (nm/\eps))$ random transformations, every hyperedge in the hypergraph would have been touched by a few of the transformations, and would look like a randomly sampled one. This will then imply the distribution of $T_{a_t,b_t}\circ\ldots\circ T_{a_1,b_1} \circ G$ and the uniform distribution from $G_{n,m,d}$ will be close. An alternative view is to think of the transformations as a Markov process, then $t = \Omega(n \log (nm/\eps))$ is the mixing time of the Markov Chain. The details of the proof are deferred to \cref{proof:random_graph} 
        \manualqed
      \end{proofsketch}
      \cref{claim:random_graph} implies that $H_t^s$ is similarly distributed as $(G^\prime, f_{G,P}(s))$ where $G^\prime \gets G_{n,m,d}$ and is unrelated to the output of the local function. Now, we would like to show that the output of the local function is statistically close to a random binary string with the same bias $\eta$.

      \begin{restatable}[Random Output]{claim}{claimrandomoutput}
          \label{claim:random_output}
         Suppose $s$ is $\eps$-fairly balanced and $G \gets G_{n,m,d}$, $m = o(n^{c/2 - 2})$, $\eps = \omega(n^{-c/4})$, $d = O(1)$ and a $c$-correlated $d$-ary predicate $P$ with bounded bias $\eta$, then the statistical distance between $f_{G,P}(s)$ and $Bern(\eta)^m$ is at most $\eps/8$.
      \end{restatable}
      \begin{proofsketch}
          Suppose that the secret $s$ is perfectly balanced, then since each output of the $d$-local function is independent conditioned on the secret, the distribution of $f_{G,P}(s)$ (for random $G$) is exactly the same as $Bern(\eta)^m$. Using the $\eps$-fairly balanced assumption on $s$, we know that the bias towards $1$ or $0$ is bounded by $1/2 + 2\log(1/\eps)/\sqrt{n}$
          
          Say the bias towards 1 is $1/2 + \alpha$. From knowing that the predicate $P$ is $c$-correlated, we can then show that $\expec{x_i \gets D}{P(x_1, \dots, x_d)} - \eta = O(\alpha^{c})$. Then, using KL divergence and Pinsker's inequality, it can be shown that one requires $\Omega(n^{c/2})$ samples to distinguish the output from $Bern(\eta)^m$. But having so many samples also means that both search and decision become efficient to solve. In particular, we use the assumption that $m = o(n^{c/2 - 2})$, $\eps = \omega(n^{-c/4})$, to show that the statistical distance is $o(\eps)$. The details are deferred to \cref{proof:random_output}.
          \manualqed
      \end{proofsketch}
      Combining both claims, we get an implication that for $t = \Omega(n \log(nm/\eps))$ with a large enough constant, $H_t^s$ has statistical distance at most $\eps/4$ from $(G, b)$ where $b \gets Bern(\eta)^m$
      \manualqed
  \end{proof}
      The next claim says that if the statistical distance is small, the decision algorithm can still be used to distinguish, even when the distribution is not exactly the same.
    \begin{claim}
        \label{claim:distinguisher_statistical_distance}
        Given $\algD$ for Decision Problem with advantage $\eps\in [0,1]$. If the distribution $null^\prime$ is taken from a distribution that has statistical distance with the null distribution of at most $\eps/k$ for constant $k > 1$, then $\algD$ can distinguish planted and $null^\prime$ with advantage at least $\Omega(\eps)$.
        \[
            \abs{ \prob{}{\algD(planted) = 1} - \prob{}{\algD(null^\prime) = 1}  } \geq \eps - \eps/k = \Omega(\eps)
        \]
    \end{claim}
    \begin{proof}
        This is because all algorithms can only distinguish $null^\prime$ and $null$ with an advantage of at most the statistical distance.
    \end{proof}
    
   \subsection{Predictor}
   
   In this subsection, we will describe the construction of a predictor that can predict the value of $s_1 \oplus s_i$ with high probability. The predictor internally constructs hybrids and decides based on the outputs of the distinguisher under these hybrids.
   
   First, we will address the technical challenge that the decision algorithm may just fail to work for some secret $s$. Since we perform a permutation on the hypergraph, the Hamming weight of the secret $s$ is fixed throughout the reduction. If we are unlucky and get a ``bad'' Hamming weight that the algorithm always fails, then recovery is impossible. Nevertheless, it can be shown that there is still a sufficiently large number of secrets (parameterised by $\eps$) which are good to distinguish.
    
  \begin{definition}[Good Secret]
    \label{def:good_secret}
      For a distinguishing algorithm $\algD$ with advantage $\eps$, define $\mathcal{G}_\algD \subseteq \Field{2}^n$ to be the set of secrets whose Hamming weight is good to distinguish. i.e.
      \[
        \mathcal{G}_\algD = \left\{s \in \Field{2}^n \ \Big|\ \prob{s^\prime \gets \F_2^n, wt(s) = wt(s^\prime)}{\algD(G, f_{G,P}(s^\prime)) = 1} - \Pr[\algD(G, b) = 1] \geq \eps/2 \right\}
      \]
      where $G \gets G_{n,m,d}, b \gets Bern(\eta)^m$, $\eta$ is the bias of $P$.
  \end{definition}

  Note that $\mathcal{G}_D$ is essentially a collection of binary strings that has some Hamming weights. If $wt(s) = wt(s^\prime)$ and $s \in \mathcal{G}_D$, then $s^\prime \in \mathcal{G}_D$.

  \begin{restatable}{claim}{goodsecretfunction}
      \label{claim:good_secret_fraction}
        Suppose $\algD$ has advantage $\eps$, then $|\mathcal{G}_\algD| \geq \eps/2\cdot (2^n)$.
  \end{restatable}
  \begin{proofsketch}
        Proven using Markov's inequality, the details are deferred to \cref{proof:good_secret_fraction}.  \end{proofsketch}
  \noindent In the next lemma, we describe the predictor algorithm for $s_i$. 
  \begin{lemma}
    \label{lemma:general_predictor} 
    Consider any $d$-ary predicate $P$ with bounded bias $\eta$, and suppose there is a polynomial-time algorithm $\algD$ that has advantage $\eps$ to solve the Decision Problem for $(P,n,m)$. Then there is a set of polynomial-time algorithms $\{\algS_2,\dots,\algS_n\}$ such that for all $\eps$-fairly balanced and good $s \in \mathcal{G}_D$, there is a pair of numbers $eq_s,neq_s\in [0,1]$ and the algorithms behave as follows for each $i\in[2,n]$:
    \begin{itemize}[itemsep=0pt]
        \item If $s_1 = s_i$, then $\Pr[\algS_i(G, f_{G,P}(s)) = 1] = eq_s$
        \item If $s_1 \neq s_i$, then $\Pr[\algS_i(G, f_{G,P}(s)) = 1] \leq neq_s$ 
    \end{itemize}
    Further, for some $t = O(n \log (mn/\eps))$, the following holds for all such $s$:
    \[
        eq_s - neq_s \geq \eps/4t
    \]
  \end{lemma}
  \begin{proof}
    Fix any $\eps$-fairly balanced and good $s \in \mathcal{G}_D$, without loss of generality, we can remove the absolute value on the advantage of $\algD$. i.e. $\Pr[\algD(planted) = 1] - \Pr[\algD(null) = 1] \geq \eps/2$. Then, the following corollary is immediate from the previous lemmas.
    \begin{corollary}
        \label{cor:hybrid_advantange} 
        There exists $t = O(n\log(nm/\eps))$ such that:
        $$\Pr_{r \gets [0, t-1]}[\algD(H_{r}^s) = 1] - \Pr_{r \gets [0, t-1]}[\algD(H_{r+1}^s) = 1] \geq\eps/4t$$
    \end{corollary}
    \begin{proof}
        Combine the statements from good secret \cref{def:good_secret}, hybrid argument \cref{claim:hybrid_argument}, terminal hybrid \cref{lemma:terminal_hybrid} and statistical distance \cref{claim:distinguisher_statistical_distance}
        \manualqed
    \end{proof}
    \noindent Now we describe the predictor algorithm $\algS_i$. On input $(G, f_{G,P}(s))$ and given blackbox access to $\algD$, it acts as follows:

    \begin{center}
      \begin{minipage}{0.85\textwidth}
        \begin{mdframed}
            \underline{\textbf{Algorithm $\algS_i(G, f_{G,P}(s))$}:}
            \begin{enumerate}[itemsep=0pt]
                \item Sample a random number $r \gets [0, t-1]$. 
                \item Sample a random permutation $\pi$ on $[n]$.
                \item Get $H_{r}^{s}$ by performing $r$ hybridization steps on $G$, i.e.\\ $H_r^{s} = (T_{a_r,b_r}\circ \ldots \circ T_{a_1,b_1}(\pi(G)), f_{G,P}(s))$, where $a_j, b_j \gets [n]$.
                \item Apply $T_{\pi(1), \pi(i)}$ to $H_r^{s}$ to obtain $H$. 
                \item Return $\algD(H)$.
            \end{enumerate}
        \end{mdframed}
      \end{minipage}
    \end{center}

    \noindent Note that the specification of $\algS_i$ does not depend on $s$ itself. From the described algorithm, we know that $\algS_i(G, f_{G,P}(s)) = D(H)$. To use \cref{cor:hybrid_advantange}, we would like to show that when $s_1 = s_i$, the distinguisher’s output on $H_r^s$ remains similar even if the input is replaced with $H$. For the case of $s_1 \neq s_i$, the comparison is with $H_{r+1}^s$ instead. 
    \begin{restatable}[Equal]{claim}{equalguess}
        If $s_1 = s_i$, then $H \approx H_r^s$.
    \end{restatable}
    \begin{proofsketch}
        When $s_1 = s_i$, this implies $\pi(s)_{\pi(1)} = \pi(s)_{\pi(i)}$, which means that the transformation $T_{\pi(1), \pi(i)}$ on the hypergraph does not affect the output of the $d$-local function. Essentially, no effective randomisation is performed, and the distribution does not look more like the null distribution. Therefore we are able to conclude that $H \approx H_r^s$. Details are deferred to \cref{proof:equal_guess}
        \manualqed
    \end{proofsketch}
    \begin{restatable}[Not Equal]{claim}{notequalguess}
        If $s_1 \neq s_i$, $\Pr[\algD(H_{r+1}^s) = 1]  \geq \Pr[\algD(H) = 1]$.
    \end{restatable}
    \begin{proofsketch}
        When $s_1 \neq s_i$, this implies $\pi(s)_{\pi(1)} \neq \pi(s)_{\pi(i)}$. Therefore, only effective randomisation is performed. The distribution $H$ looks more like the null distribution than $H_{r+1}^s$, because the last transformation $T_{a,b}$ on $H_{r+1}^s$ could be effective or not, depending on the choice of $a,b$. Since the distinguisher tends to return $1$ less often when given the null distribution, when $H$ is given as an input, it should return $1$ less often than $H_{r+1}^s$. See \cref{proof:notequal_guess} for details.
        \manualqed
    \end{proofsketch}
    \noindent As noted above, $\pr{\algS_i(G, f_{G,P}(s))=1} = \pr{D(H)=1}$. Set $eq_s = \Pr[\algD(H_{r}^s) = 1]$ and $neq_s = \Pr[\algD(H_{r+1}^s) = 1]$. Combining \cref{cor:hybrid_advantange} and the two claims above then proves the claimed behaviour of the $\algS_i$'s, and also yields $eq_s - neq_s \geq \eps/4t$.
    \manualqed
    \end{proof}

    \subsection{Amplification}
    
    Now we will show how to amplify the predictor's advantage for each $i \in [2,n]$, and eventually recover $s$. To use the predictor, we first need to guess the value of $s_1$. Since there are only 2 possible values, $0$ and $1$, we can just try both values. This will ultimately give two candidates as the solution to the problem. The correct one can then be verified by re-evaluating the $d$-local function on the candidates.

    \medskip

    Next, as the $\algS_i$'s only guarantee a gap in the response based on the relationship of $s_1$ and $s_i$, to correctly identify the relationship with high confidence, we will need to estimate the response of $\algS_i$ when $s_1 = s_i$. More precisely, the following value needs to be estimated
    \[
        eq_s = \Pr_r[D(H_{r}^s) = 1]
    \]
    Furthermore, the value $eq_s$ could vary based on the value of $s$. Fortunately, since for secrets $s$ with the same Hamming weight, we know that their $eq_s$ is the same, as in our construction of the hybrids the first step is to effectively permute the secret. We just have to perform this estimation for one secret for each possible Hamming weight.
    
    The estimation can be done by locally generating a secret of a given Hamming weight and a random problem instance, then applying hybridisation and sending it to the distinguisher $\algD$. With $O((t/\eps)^4\log (n/\eps))$ number of trials, by Hoeffding's inequality, the average will converge to within $o(\eps/t)$ of the true value of $eq_s$, with failure probability $o(\eps)$. Once we have these estimates, we run the computations below with each of them until the correct one is used and the secret $s$ is found. 
    
    \medskip
    Given the value of $eq_s$, fix some $i\in[2,n]$. We can amplify the prediction probability by rerunning the predictor $\algS_i(G,f_{G,P}(s))$ with a fresh new problem instance of the same secret $s$, with an independent new $G$. Suppose $l$ repetitions are performed, take the sum of the outputs of $\algS_i$ and put a threshold at $l(eq_s - \eps/8t)$. If it is greater than that, return $s_1$ as the guess for $s_i$; otherwise, return the flip of $s_1$. Formally, let $X$ be the random variable of the sum of the output of the distinguisher. \\\\
    \noindent If $s_1 \neq s_i$, $\exp{X} \leq l(eq_s -\eps/4t)$, by Hoeffding's bound
    \[
        \Pr[X \geq l(eq_s -\eps/8t)] \leq \exponential(-O(l\eps^2/t^2))
    \]
    If $s_1 = s_i$,  $\exp{X} = l\cdot eq_s$, by Hoeffding's bound
    \[
        \Pr[X \leq l(eq_s - \eps/8t)] \leq \exponential(-O(l\eps^2/t^2))
    \]
    Therefore, the failure probability is at most $\exponential(-O(l\eps^2/t^2))$. 
    
    Now that we have a predictor of $s_i$ with high probability, use the same set of problem instances to learn other $s_j$ for $j \in [2,n]$. Performing a union bound over all $i$, we set $l = \Theta(t^2\log (n/\eps)/\eps^2)$ for a failure probability of at most $\eps/4$ in guessing correctly whether $s_i=s_1$ for all $i\in[2,n]$. 

    \medskip
    Finally, we can now prove our main theorem \cref{thm:general}. Since we assumed that $d = O(1)$, the $d$-ary $c$-correlated predicate $P$ must have bounded bias (depending on $d$), and we can assume $m = o(n^{c/2 - 2})$ and $\eps = w(n^{-c/4})$. The recovery of the secret in total costs $O((n^2/\eps^2)\log^3(n/\eps)) \times m$ samples and runs in time polynomial of $n,m$ and $T$ (runtime of decision algorithm). The success probability is determined by the number of good secrets and whether the secret is $\eps$-fairly balanced, as the above algorithm works if both these conditions are met. The probability that a randomly selected secret is not good is $1 - \eps/2$ (\cref{claim:good_secret_fraction}), the probability that a randomly selected secret is not fairly balanced is $o(\eps)$ (\cref{rem:fairly_balance_prob}). By union bound, the probability that a randomly selected secret is both good and fairly balanced and the algorithm succeeds is $\Omega(\eps)$. Finally, if the original distinguisher runs in polynomial time, then all of the above can be done in polynomial time as well. This proves \cref{thm:general}.


\section{Generalization}
\label{sec:generalization}

In this section, we will describe several interesting generalisations of our technique to a larger family of problems. Throughout, we say that a predicate has \emph{bounded bias} if its bias is bounded away from $0$ and $1$ (see \cref{def:bounded_bias}).

\subsection{Non-Constant Sparsity}
\label{sec:non-constant-sparsity}

\cref{thm:general} has the assumption that the sparsity $d$ must be a constant. A natural question is whether the same result holds for larger non-constant sparsity, say $d = \log n$, and to what extent the techniques fail. The parts of the proof of \cref{thm:general} that requires constant sparsity are \cref{claim:random_output} (that the output distribution of the function is close to Bernoulli), and the  sample threshold lemma (\cref{thm:threshold}, which gives algorithms for large enough $m$, letting us assume in our proof that $m = o(n^{c/2-2})$ and $\eps = \omega(n^{-c/4})$). Unfortunately, the latter lemma might well not be true when $d = \omega(1)$. 

Regardless, we can still generalise the relationship between $m$ and $\eps$ for which our reduction applies. In particular, we generalise \cref{claim:random_output} so that it allows non-constant locality. The requirement that we will need to add to allow our reduction to work for $d = \polylog(n) = O(\log^r(n))$ for any constant $r > 0$ is:
\begin{equation}
    \label{eq:large_sparsity}
    (m/n^c) \cdot (\log^{r}(n)\log(1/\eps))^{2c} = o(\eps^2) \text{ and } \log^r(n)\log(1/\eps) = o(\sqrt{n})
\end{equation}
Although these conditions may appear complicated, they are easily satisfied when $\eps = w(n^{-c/4})$, $m = o(n^{c/2})$ (a stronger condition). The generalised theorem is as follows.

\begin{theorem}
  For $d = O(\log^r n)$ with constant $r > 0$ and $c\leq d$, consider a $c$-correlated $d$-ary predicate $P$ with bounded bias. Suppose there is a polynomial-time algorithm that has advantage $\eps \in [0,1]$ in solving the Decision Problem for $(P,n,m)$ for some $m = poly(n)$ such that \cref{eq:large_sparsity} is satisfied. Then there is a polynomial-time algorithm for the Search Problem for $(P,n,\ell m)$ that has success probability $\Omega(\eps)$, for some $\ell = \Theta((n/\eps)^2\log^3 (n/\eps))$.
\end{theorem}

\begin{proofsketch}
    The proof is the same as that of \cref{thm:general}, except for the proof of the Terminal Hybrid \cref{lemma:terminal_hybrid}, where we need to re-state \cref{claim:random_output} to account for non-constant locality, as follows.

    \begin{claim}
        \label{claim:random_output_gen}
        Suppose $s$ is $\eps$-fairly balanced and $G \gets G_{n,m,d}$, with $d = O(\log^r n)$, a $c$-correlated $d$-ary predicate $P$ with bounded bias $\eta$ and \cref{eq:large_sparsity} is satisfied, then the statistical distance between $f_{G,P}(s)$ and $Bern(\eta)^m$ is at most $\eps/8$
    \end{claim}

    \begin{proofsketch}
        Most of the proof remains the same as that of \cref{claim:random_output}; the first change that we have to make is that the bound we get on the bias of $P$ when given non-uniform inputs is slightly different, requiring an additional $\log^{cr}(n)$ factor.
        \[
            \expec{x_i \gets D}{P(x_1, \dots, x_d)} - \eta  
            = O\left(\sum_{k = c}^{d} \left(\dfrac{ed\log (1/\eps)}{k\sqrt{n}}\right)^k\right)
            = O\left(\dfrac{(\log^r(n)\log(1/\eps))^{c}}{\sqrt{n}^c}\right)
        \]
        Where the last equality is from the assumption that $\log^r(n)\log(1/\eps) = o(\sqrt{n})$. \\\\
        The second change is in the final part where we show the statistical distance between $f_{G,P}(s)$ and $Bern(\eta)^m$ is $o(\eps)$, where we will use the assumption that $(m/n^c) \cdot (\log^{r}(n)\log(1/\eps))^{2c} = o(\eps^2)$ as follows.
        \[
            O(\sqrt{m\alpha^{2c}}) = O\left(\sqrt{m \times\dfrac{(\log^r(n)\log(1/\eps))^{2c}}{n^c}}\right) = o(\eps)
        \]
    \manualqed
    \end{proofsketch}
    \manualqed
\end{proofsketch}

\subsection{Distinct values in the hyperedges}
\label{sec:distinct-hyperedge}

Another common definition for $d$-local functions is to have the requirement that in each edge of the hypergraph, every vertex is distinct -- that is, for any output bit, the same bit is not given twice to the predicate as input when computing it. Our technique still works in this model, with a slightly larger number of hybrids $t$ and with $d = O(\log^r n)$. There are two claims in the proof that are affected by this change, which are \cref{claim:random_graph} and \cref{claim:random_output}, both about the uniformity of the terminal hybrid. 

For \cref{claim:random_graph}, it could be that for the transformation $T_{a,b}$ that is applied, there are hyperedges that contain both $a$ and $b$, which then cannot be replaced independently. The easiest modification to do here is to not perform any changes in this event. Formally, the transformation $T_{a,b}$ is modified as follows: for a hypergraph $G = (S_i)_{i \in [m]}$, each $S_i = (j_1, \dots, j_d)$ is transformed to $S_i^\prime = (j_1^\prime, \dots, j_d^\prime)$ independently where
      \[
            \Pr[j_k^\prime = j_k] = 1 \quad\text{ if } j_k \notin \{a, b\} \text{ or } \{a, b\} \subseteq \{j_1, \dots, j_d\}\\
      \]
      \[
            \Pr[j_k^\prime = a] = \dfrac{1}{2},\ \Pr[j_k^\prime = b] = \dfrac{1}{2}  \quad\text{ otherwise }
      \]      
Then, for each edge, on expectation, there will be around $t \times (n^2 - d^2)/n^2$ transformations that have $\{a, b\} \not\subseteq \{j_1, \dots, j_d\}$. Call these transformations \emph{effective}. The expected number of effective transformations $t^\prime$ is asymptotically the same $t$ when $d$ is small (say $d = O(\log^r n)$). 
\[
    \expec{a, b \gets [n]}{t^\prime} = t \times (n^2 - d^2)/n^2 = \Theta(t)
\]
Since each choice of $a, b$ is independent, using concentration bounds, using $10\cdot t$ transformations will ensure that there will be at least $t$ effective transformations for every edge with probability at least $1 - \exponential(-\Omega(t))$, which is also $1 - o(\eps)$.

The other issue is in \cref{claim:random_output}, where after an index in $s$ is selected, it cannot be chosen again. Suppose the input contains more ones than zeros. In the worst case, the previously selected inputs are all zeros, which only increases the fraction of ones among the remaining inputs. Nevertheless, as long as $d$ is small enough, the bias will still be bounded by $1/2 + O(\log(1/\eps)/\sqrt{n})$ with a different constant factor. This is still considered as $\poly(\eps)$-fairly balanced secret, and would not be an issue asymptotically. 

Define the Distinct Decision (respectively Search) Problem as the Decision (respectively Search) Problem, but with the hypergraph restricted to having distinct vertices in the hyperedges.

\begin{theorem}
  For $d = O(\log^r n)$ with constant $r > 0$ and $c\leq d$, consider a $c$-correlated $d$-ary predicate $P$ with bounded bias. Suppose there is a polynomial-time algorithm that has advantage $\eps \in [0,1]$ in solving the Distinct Decision Problem for $(P,n,m)$ for some $m = poly(n)$ such that \cref{eq:large_sparsity} is satisfied. Then there is a polynomial-time algorithm for the Distinct Search Problem for $(P,n,\ell m)$ that has success probability $\Omega(\eps)$, for some $\ell = \Theta((n/\eps)^2\log^3 (n/\eps))$.
\end{theorem}

\begin{proofsketch}
    The proof is almost the same as that of \cref{thm:general}, except that first we replace $G_{n,m,d}$ to be the set of all hypergraphs having distinct vertices in the hyperedge. Next, we modify our transformation $T_{a,b}$ to not perform any swap when both $a,b$ are in the hyperedge. It is easy to see that the uniform distribution over $G_{n,m,d}$ is still stable under this modified transformation. Then, we increase the number of transformations by a constant factor, say $10\cdot t$, to still obtain \cref{claim:random_graph} with probability $1 - \exponential(\Omega(t)) \geq 1 - o(\eps)$.
    
    Aside from the changes required to account for larger $d$ (see \cref{claim:random_output_gen}), we will also need to modify the proof of the bias to account for distinct values. (Refer to \cref{proof:random_output})
    \[
        \expec{x_i \gets D}{P(x_1, \dots, x_d)} - \eta  = -\dfrac{1}{2}\sum_{S \subseteq [d], |S| \geq c} \hat{P^\prime}(S)\exp{\prod_{i \in S}y_i}
    \]
    We cannot expand the expectation as a product of expectations because they are not independent. However, we can still do conditional expansion of the expectation. $\exp{\prod_{i \in S}y_i} = \exp{y_j \exp{\prod_{i \in S, i \neq j} y_i\ |\ y_j}}$. 
    
    Conditioned on the previous choices, the bias of the distribution does not change significantly, $\exp{y_1 | y_{2}, \dots, y_{d}}$ $=$ $\dfrac{1}{2} + O(\dfrac{\log(1/\eps)}{\sqrt{n}})$. Therefore, we still achieve $\expec{x_i \gets D}{P(x)} - \eta = O(\dfrac{\log^c (1/\eps)}{\sqrt{n}^c})$. The rest of the proof follows.
    \manualqed
\end{proofsketch}

\subsection{Noisy Predicate}
\label{sec:noisy-predicate}

Motivated by the LPN problem, we also apply our reduction to the case of the predicate $P$ being noisy. Namely, the predicate $P$ is defined as
\[
    P(x_1, \dots, x_d) = R(x_1, \dots, x_d) + e
\]
for some deterministic function $R: \Field{2}^d \to \Field{2}$, where $e$ is an independent Bernoulli random variable. 

For a noisy predicate $P$, one affected part in our proof is that we cannot use the Fourier Transformation as is in the proof of \cref{claim:random_output}. Nevertheless, we still have the analogous claim on the deterministic predicate $R$, and it is easy to see that the error would only decrease the statistical distance between $f_{G,R}(s)$ and $Bern(\exp{R})^m$. 
\[
    \Delta(f_{G,P}(s), Bern(\exp{P})^m) \leq \Delta(f_{G,R}(s), Bern(\exp{R})^m) 
\]

This is due to the fact that the noise distribution added to both distributions is the same. However, since the bias of $P$ might not be the same as the bias of $R$, we do need $R$ to have bias bounded away from $0$ and $1$ for this to work. If $d = O(1)$, then any non-constant $R$ will always have bounded bias.

The other part that is affected is the verification of the secret. Since noise is added, and our algorithm returns two secret candidates with $s_1 = 0$ or $s_1 = 1$ for each $\eps$-fairly balanced Hamming weight, it could be difficult to identify which is the correct solution. One could just return the answer that is more likely to be correct in that case.

\begin{theorem}
  For $d = O(\log^r n)$ with constant $r > 0$ and $c\leq d$, consider a $c$-correlated $d$-ary noisy predicate $P$ with bounded bias. Suppose there is a polynomial-time algorithm that has advantage $\eps \in [0,1]$ in solving the Decision Problem for $(P,n,m)$ for some $m = poly(n)$ such that \cref{eq:large_sparsity} is satisfied. Then, for some $\ell = \Theta((n/\eps)^2\log^3 (n/\eps))$, there is a polynomial-time algorithm for the Search Problem for $(P,n,\ell m)$ that, with probability $\Omega(\eps)$, returns a set of secrets of size at most $2n$ that contains the solution.
\end{theorem}

\begin{proof}
    The proof is the same as \cref{thm:general}, except at the proof of \cref{claim:random_output} (refer to \cref{proof:random_output}), we will first perform Fourier transformation on the deterministic portion of $P$, say $R$. Because the noise distribution added to both distributions is the same, we claim that noise will only result in a lower statistical distance. More formally, suppose $\eta = \exp{R}$ and $\beta \in [0,1/2]$ is the parameter for noise, we will need to show
    \begin{align*}
        &\quad D_{KL}(Bern(\eta + \alpha) \oplus Bern(\beta)||Bern(\eta) \oplus Bern(\beta))\\
        &\leq D_{KL}(Bern(\eta + \alpha) ||Bern(\eta))
    \end{align*}
    The difference of probability of $Bern(\eta + \alpha) \oplus Bern(\beta) = 1 $ and probability of $Bern(\eta ) \oplus Bern(\beta) = 1$ is $\alpha(1 - 2\beta)$, which is smaller than $\alpha$ (the original difference). Therefore, the distribution is closer, which is why the KL divergence is smaller. 

    On the verification part, since we might not be able to perfectly identify which is the correct solution, we just return all candidate solutions in a set $\mathcal{S}$. For each Hamming weight, we will obtain 2 solutions, therefore the size of $\mathcal{S}$ is at most $2n$.
    \manualqed
\end{proof}

\section{Deferred Proofs}
\label{sec:deferred}

\subsection{Random Graph}
\claimrandomgraph*
      \begin{proof}
      \label{proof:random_graph}
          To show the claim, it suffices to show that for a hyperedge $S_i = (j_1, \dots, j_d)$, after the transformations, say $S^\prime_i = (j^\prime_1, \dots, j^\prime_d)$,  each value $j^\prime_k$ is independently distributed like randomly sampled from $[n]$. 

          Define a distribution of values as a vector $D = (x_1, \dots, x_n) \in [0,1]^n$, such that $\sum x_i = 1$. The value $x_k$ indicates the probability that sampling from $D$ gives the value $k$. Define $D(i) = (x_1(i), \dots, x_d(i))$ as the distribution of the value after applying $i$-th transformation to $D$. 

          We represent the distribution of a value in the hyperedge with $D$. Suppose we are working on vertex $j_1$ in a hyperedge $S_i$. Initially, the distribution $D(0)$ looks like $(0, 0, \dots, 1, \dots, 0)$. Basically, only $x_{j_1} = 1$ and the rest of the values are $0$. Sampling from $D(0)$ gives $j_1$ with probability $1$. Observe that when $T_{a,b}$ is applied to a hypergraph, its effect on $D(i)$ is that $x_a(i)$ and $x_b(i)$ are averaged, while the rest remains unchanged. Writing it down, 
          \[
            x_a(i+1), x_b(i+1) = (x_a(i) + x_b(i))/2
          \]
          This is because $T_{a,b}$ only changes the value of a vertex in a hypergraph if it is either $a$ or $b$. Otherwise, it remains unchanged. Our goal is to prove that the random averaging process converges to a uniform distribution; namely, when $t = \Omega(n \log (n/\eps))$, $\sum_{k \in [n]}(\abs{x_k(t) - 1/n})$ becomes small. This then somewhat resembles the randomised gossip process, which we can then use a similar approach from \cite{BGPS06} to show convergence. To prove that, we use the $L_2$ deviation function $V$ defined as 
          \[
            V(i) = \sum_{k \in [n]} (x_k(i) - \dfrac{1}{n})^2
          \]
          We claim that $0 \leq V(i) \leq 1$ by showing that $V(i)$ is a non-increasing function as $i$ increases and $V(0) \leq 1$. First on $V(0) \leq 1$, writing it down
          \[
            V(0) = \dfrac{(n-1)}{n^2} + (\dfrac{n-1}{n})^2 = \dfrac{n-1}{n} \leq 1
          \]
          Next on non-increasing, with a transformation of $T_{a,b}$, the initial contribution from $x_a$ and $x_b$ to the deviation is $(x_a - 1/n)^2 + (x_b - 1/n)^2$. Let $u = x_a - 1/n, v = x_b - 1/n$. After the transformation, the values are averaged and the contribution to $V(i+1)$ becomes 
          \[ 2(\dfrac{x_a + x_b}{2} - \dfrac{1}{n})^2 = 2(\dfrac{u + v}{2})^2 = \dfrac{(u+v)^2}{2}
          \]
          The difference in the value of $V(i)$ and $V(i+1)$ is then 
          \[
            V(i) - V(i+1) = u^2 + v^2 - \dfrac{(u+v)^2}{2} = \dfrac{(u - v)^2}{2}  = \dfrac{(x_a - x_b)^2}{2}\geq 0
          \]
          Therefore, $V(i)$ is a non-increasing function. The next step is to show that $V(t)$ decays quickly to $0$ when we apply random transformations. The expected difference after applying a transformation is
          \begin{align*}
              \expec{a, b \gets [n]}{(x_a - x_b)^2/2} &= \dfrac{1}{2}\expec{a, b \gets [n]}{x_a^2 - 2x_ax_b + x_b^2} \\
              &= \dfrac{1}{n}\sum_{a \in [n]} x_a^2 - \dfrac{1}{n^2} = \dfrac{1}{n} \sum_{a \in [n]} (x_a - \dfrac{1}{n})^2 = \dfrac{V(i)}{n}
          \end{align*}
          Where the second equality is due to $\exp{x_a^2} = \exp{x_b^2}$ and $\exp{x_a} = \exp{x_b} = 1/n$. The third equality is due to the definition of variance $\frac{1}{n}\sum_{a \in [n] } x_a^2 - \frac{1}{n^2} = \exp{x_a^2} - \exp{x_a}^2 = \exp{(x_a - \frac{1}{n})^2}$.\\\\
          Therefore,
          \[
            \exp{V(i+1)\ |\ V(i) = v_i} = v_i - \exp{(x_a - x_b)^2/2 \ |\ V(i) = v_i} = v_i - \dfrac{v_i}{n} = (1 - \dfrac{1}{n}) v_i
          \]
          So by law of total expectation,
          \[
              \exp{V(i+1)} = \int \left(\exp{V(i+1)\ |\ V(i) = x}\pr{V(i) = x}\right) dx = (1 - \dfrac{1}{n})\exp{V(i)}
          \]
          Starting with $V(0) \leq 1$, this implies $\exp{V(t)} \leq (1 - 1/n)^t \leq \exponential({-t/n})$. By Markov inequality,
          \begin{equation}
            \label{eq:markov}
              \Pr_{a_{j}, b_{j} \gets [n]}[V(t) \geq (md/\eps^2)\exponential({-t/n})] \leq \eps^2/md
          \end{equation}
          When $V(t) \leq (md/\eps^2)\exponential({-t/n})$, we use Cauchy–Schwarz inequality to bound the statistical distance between $V(t)$ and the uniform distribution
          \[
            \sum_{k \in [n]}(\abs{x_k(t) - 1/n}) \leq \sqrt{nV(t)} \leq \dfrac{\sqrt{nmd}}{\eps}\cdot \exponential({-t/2n}) 
          \]
            
          Each vertex in the hyperedge has its own distribution $D$. Note that their distributions for different vertices are not independent due to the choice of $a_j, b_j$. It could also not be identical due to different initial states. Nevertheless, each vertex is sampled independently from its own distribution $D$.

          We say that a distribution of a vertex is bad if the $(a_j,b_j)$'s are chosen in such a way that $V(t)$ for that vertex is larger than $z = (md/\eps^2)\exponential(-t/n)$. The probability that the $D$ is bad is at most $\eps^2/md$ (\cref{eq:markov}). By union bound, across the $md$ vertices in all the hyperedges, there is a distribution that is bad with probability at most $\eps^2$. In the case of all distributions being good, the distance from uniform for one of the vertices is at most $\sqrt{nz}$. Therefore, the overall statistical distance between the product distribution and a randomly sampled hypergraph can be upperbounded by $\eps^2 + md \cdot \sqrt{nz}$
          
          As $m = poly(n)$ and $d = O(1)$, set $t = 8n \log (mdn/\eps) = \Theta(n \log (nm/\eps))$ to have the overall statistical distance at most $2\eps^2$. We can then make the statistical distance smaller than $\eps/8$ by adjusting the constant factor.
          \manualqed
      \end{proof}

\subsection{Random Output}

\claimrandomoutput*
\begin{proof}
    \label{proof:random_output} 
    We start the proof by introducing the Fourier transformation and explaining the connection between $c$-correlation and the Fourier coefficient.

    \begin{remark}[Fourier Transformation] \emph{\cite[Chapter 1]{odonnell-book}}
    \label{rem:fourier}
    Let a $d$-ary predicate $P$, convert the working field from $\Field{2}$ to $\mathbb{R}$ by mapping $0, 1$ in $\Field{2}$ to $1, -1$ in $\mathbb{R}$. So we have $P: \{-1, 1\}^d \to \{-1, 1\}$. It can then be uniquely expressed as a multilinear polynomial
    \[
        P(x_1, \dots ,x_d) = \sum_{S \subseteq [d]} \hat{P}(S)\prod_{i \in S} x_i
    \]
    where the Fourier coefficient $\hat{P}(S) = \exp{P(x_1, \dots, x_d) \prod_{i \in S}x_i}$
    \end{remark}

    When a predicate is $c$-correlated, by minimality of $c$, it also means the Fourier coefficient for all non-empty subsets of size less than $c$ is zero. Now, we proceed to the main argument.

    Since $s$ is $\eps$-fairly balanced, $s$ has Hamming weight bounded by $[n/2 - w, n/2 + w]$ where $w = 2\sqrt{n}\log (1/\eps)$. Without loss of generality, assume that $s$ has at least as many ones as zeros. Let $D$ be the distribution of randomly sampling an element in the binary string $s$. Then, $D$ is a Bernoulli distribution that is only slightly biased. 
    \[
        \exp{D} - 1/2 = \dfrac{2\log (1/\eps)}{\sqrt{n}}
    \]
    Let $P^\prime, D^\prime$ be the analog of $P, D$ with output in $\{-1, 1\}$ and working field of $\mathbb{R}$. Assuming that this good event happens, from the Fourier expansion of $c$-correlated predicate $P^\prime$ (\cref{rem:fourier}), we have
    \begin{align*}
         \expec{x_i \gets D}{P(x_1, \dots, x_d)} - \eta &= -\dfrac{1}{2}\sum_{S \subseteq [d], |S| \geq c} \hat{P^\prime}(S)\prod_{i \in S} \exp{y_i} \\
        = O\left(\sum_{k = c}^{d} {\binom{d}{k}} \cdot \dfrac{\log^k (1/\eps)}{\sqrt{n}^k}\right) &= O\left(\sum_{k = c}^{d} (\dfrac{ed\log (1/\eps)}{k\sqrt{n}})^k\right) = O(\dfrac{\log^c (1/\eps)}{\sqrt{n}^c})
    \end{align*}
    Where the last equality is due to $d = O(1)$. It is clear to see that if $D$ is closer to $Bern(1/2)$, then the distribution of $P$ when its inputs are sampled from $D$ will also be closer to $Bern(\eta)$. 
    
    More precisely, let $\alpha = h_1\log (1/\eps)/\sqrt{n}$ for some constant $h_1 > 0$, and $D = Bern(1/2 + \alpha)$, then the output distribution of $d$-local function $f_{P,G}(s)$ (over random $G$) is $Bern(\eta + h_2\alpha^c)^m$ for some constant $h_2 > 0$ (the output bits are independent conditioned on the secret $s$). So as $\alpha$ goes smaller or $c$ goes larger, it is closer to $Bern(\eta)^m$.
    
    To quantitatively understand the distance between $Bern(\eta + \alpha^c)^m$ and $Bern(\eta)^m$. Use Pinsker's inequality that relates the statistical distance of two distributions with the KL divergence,
    \begin{align*}
        \Delta(Bern(\eta + \alpha^c)^m, Bern(\eta)^m) &\leq \sqrt{D_{KL}(Bern(\eta + \alpha^c)^m|| Bern(\eta)^m)} \\
        &= \sqrt{m \times D_{KL}(Bern(\eta + \alpha^c)|| Bern(\eta))}
    \end{align*}
    An upper bound on the KL divergence can be achieved by performing Taylor expansion on the KL divergence \cite[Problem 11.2]{thomas-book}, which gives
    \[
    D_{KL}(Bern(\eta + \alpha^c)||Bern(\eta)) = O\left(\alpha^{2c}/\eta(1-\eta)\right) = O(\alpha^{2c})
    \]
    The final inequality is due to $\eta$ being a bounded bias.
    Finally, the statistical distance is then
    \begin{align*}
        O(\sqrt{m\alpha^{2c}}) &= O\left(\sqrt{m \times\dfrac{\log^{2c}(1/\eps)}{n^c}}\right) = o(\dfrac{(c/4)^c\log^c(n)}{n^{c/4+1}}) = o(\eps)
    \end{align*}
    Where the second equality is due to $m = o(n^{c/2-2})$ and $\eps = \omega(n^{-c/4})$. The last inequality is from $\eps = \omega(n^{-c/4})$. Since the statistical distance is $o(\eps)$, we can then make the statistical distance smaller than $\eps/8$ by adjusting the constant factor.
    \manualqed
\end{proof}

\subsection{Good Secret}

\goodsecretfunction*
  \begin{proof}
    \label{proof:good_secret_fraction} 
    Suppose $\eta$ is the bias of $P$, let
    \[
     p^s_0 = \prob{\substack{G \gets G_{n,m,d}\\s^\prime \gets \F_2^n, wt(s) = wt(s^\prime)}}{\algD(G, f_{G,P}(s^\prime)) = 1},\quad p^s_1 = \prob{\substack{G \gets G_{n,m,d}\\ b \gets {Bern(\eta)^m}}}{\algD(G, b) = 1}
    \]
    Then,
    \begin{align*}
        \expec{s \gets \Field{2}^n}{p_0^s} = \sum_{s \in \Field{2}^n} \dfrac{1}{2^n}p_0^s 
        &=  \sum_{k = 0}^{n} \sum_{\substack{s \in \Field{2}^n\\wt(s) = k}} \dfrac{1}{2^n} \cdot \prob{\substack{G \gets G_{n,m,d}}}{\algD(G, f_{G,P}(s)) = 1}\\
        &= \prob{\substack{G \gets G_{n,m,d}\\s^\prime \gets \F_2^n}}{\algD(G, f_{G,P}(s')) = 1}
    \end{align*}
    So, $\abs{\exp{p_0^s} - \exp{p_1^s}}  \geq \eps$. Using linearity of expectation and without loss of generality, $\exp{p_0^s - p_1^s} \geq \eps$. Let $q_s = 1 - (p_0^s - p_1^s)$, then $q_s \in [0, 2]$ and $\exp{q_s} \leq 1 - \eps$. Using Markov's inequality
    \[
        \prob{s \gets \Field{2}^n}{p_0^s - p_1^s < \eps/2} = \prob{s \gets \Field{2}^n}{q_s > 1 - \eps/2} \leq \dfrac{1 - \eps}{1 - \eps/2} = 1 - \dfrac{\eps/2}{1 - \eps/2}
    \]
    Which implies
    \[
        \prob{s \gets \Field{2}^n}{p_0^s - p_1^s \geq \eps/2} \geq \dfrac{\eps/2}{1 - \eps/2} \geq \eps/2
    \]
    Therefore there should be at least $\eps/2$ fraction of secret $s \in \Field{2}^n$ such that $s \in \mathcal{G}_\algD$, which implies $|\mathcal{G}_\algD| \geq (\eps/2)\cdot 2^n$
    \manualqed
  \end{proof}

\subsection{Equal}
\equalguess*

\begin{proof}
    \label{proof:equal_guess}
            If $s_1 = s_i$, intuitively, $H$ is still $H_{r}^s$ because the secret bits are the same. The hybridisation is not really performing any randomisation as the output remains consistent. To prove this more formally, for technical reasons, we will need to define the inverse of the transformation $T_{a,b}$. However, since $T_{a,b}$ is a randomised function, whose inverse is not defined, derandomisation is then necessary.

        \begin{definition}[Derandomized Transformation]
            Define a \emph{deterministic transformation} : $T^\prime_{a,b} : \Field{2}^{md} \times G_{n,m,d} \to G_{n,m,d}$. On input of a vector $v \in \Field{2}^{md}$ and a hypergraph $G$, each of the hyperedges $S_i = (j_{1}, \dots, j_d)$ is transformed to $S_i^\prime = (j_1^\prime, \dots, j_d^\prime)$ as follows:
            \[
                j_k^\prime  = 
                \begin{cases}
                    j_k &\text{if } j_k \notin \{a, b\} \text{ or } v_{i\cdot d + k} = 1\\
                    b    &\text{if } j_k = a \text{ and } v_{i\cdot d + k} = 0\\
                    a    &\text{if } j_k = b \text{ and } v_{i\cdot d + k} = 0
                \end{cases}
            \]
        \end{definition}
        \noindent The randomised transformations from our definitions of hybrids can be formed by sampling a vector $v$ uniformly from $\Field{2}^{m \times d}$ and applying the above deterministic transformation with that vector. It is also easy to see that given the vector $v$, the transformation is reversible. Therefore, we define the inverse of $(T^\prime_{a,b})^{-1}: \Field{2}^{m \times d} \times G_{n,m,d} \to G_{n,m,d}$ to reverse the process of $T^\prime$. Now, we can show an alternative view on the hybrids
        \begin{claim}[Alternative View on Hybrid]
            $H_r^s \approx (K, f_{K^\prime, P}(\pi(s)))$ where $K \gets G_{n,m,d}$, $\pi$ is a random permutation on $[n]$ and $K^\prime = (T^\prime_{a_{1},b_{1}})^{-1}(v_1, \cdot)\circ \ldots \circ (T^\prime_{a_r,b_r})^{-1}(v_r, K)$ where $v_i \gets \Field{2}^{md}$ and $a_j,b_j \gets [n]$
        \end{claim}
        \begin{proof}
            We previously defined hybrid as $H_r^s = (T_{a_r,b_r} \circ \ldots \circ T_{a_1,b_1}(\pi(G)), f_{G,P}(s))$ with $G \gets G_{n,m,d}$ (\cref{def:hybrid}). First, rewrite it with the derandomised transformation
            \[
                H_r^s = (T^\prime_{a_r,b_r}(v_r, \cdot)\circ \ldots \circ T_{a_1,b_1}(v_1, (\pi(G)), f_{G,P}(s))
            \]
            where $v_i \gets \Field{2}^{md}$. Also, for any permutation $\pi$ on $[n]$, $f_{G,P}(s) = f_{\pi(G), P}(\pi(s))$. Therefore, if we let $K = T^\prime_{a_r,b_r}(v_r, \cdot)\circ \ldots \circ T_{a_1,b_1}(v_1, (\pi(G))$, then
            \[
                K^\prime = \pi(G) = (T^\prime_{a_{1},b_{1}})^{-1}(v_1, \cdot)\circ \ldots \circ (T^\prime_{a_r,b_r})^{-1}(v_r, K)
            \]
            So $H_r^s = (K, f_{G,P}(s)) = (K, f_{\pi(G),P}(\pi(s))) = (K, f_{K^\prime, P}(\pi(s)))$. Since $G \gets G_{n,m,d}$ and the uniform distribution is stable under the transformations, so $K \approx G_{n,m,d}$.
            \manualqed
        \end{proof}
        In our case of $H$, we additionally apply $T_{\pi(1), \pi(i)}$. Using the alternative view on hybrid, $H_r^s = (K, f_{K^{\prime},P}(\pi(s)))$, we have:
        \[ 
        H = (K, f_{K^{\prime\prime},P}(\pi(s)))
        \]
        where 
        \[ 
        K^{\prime\prime} =  (T^\prime_{a_{1},b_{1}})^{-1}(v_1, \cdot)\circ \ldots \circ (T^\prime_{a_r,b_r})^{-1}(v_r, \cdot) \circ (T^\prime_{\pi(1),\pi(i)})^{-1}(v, K)  
        \]
        where $v_i \gets \Field{2}^{md}$. But since $s_1 = s_i$, so $\pi(s)_{\pi(1)} = \pi(s)_{\pi(i)}$. This implies that applying $(T^\prime_{a_r,b_r})^{-1}$ does not change the value of the $d$-local function. i.e.
        \[
            f_{K^{\prime\prime}, P}(\pi(s)) = f_{K^\prime, P}(\pi(s))
        \]
        Which proves that $H \approx H_r^s$.
        \manualqed
        \end{proof}

\subsection{Not Equal}

\notequalguess*
    \begin{proof}
        \label{proof:notequal_guess}
            Observe that $H_{r+1}^{s}$ is a mixture distribution of $H_{r}^{s}$ and $H$. In the proof of \cref{proof:equal_guess}, we have already shown that, suppose $\pi(s)_a = \pi(s)_b$ and $T_{a,b}$ is the last transformation applied to the hypergraph for $H^s_{r+1}$, then no randomisation is applied and it is distributed as $H_r^s$. Let $\pi(s) = s^\prime$ and $A$ be the event that  $s^\prime_a = s^\prime_b$, and $B$ be the event that $s^\prime_a \neq s^\prime_b$. We can say:
            \begin{equation}
                \label{eq:mix_distribution}
                \Pr[H_{r+1}^{s} = h] = \Pr_{a,b}[A]\Pr[H_r^{s} = h] + \Pr_{a,b}[B]\Pr[H = h]
            \end{equation}
            This is because in the last transformation that we apply to $H_r^s$ to get $H$, our assumption on $s_1 \neq s_i$ ensures that $T_{\pi(1), \pi(i)}$ always has $s^\prime_{\pi(1)} \neq s^\prime_{\pi(i)}$. Furthermore, the randomness of $\pi$ ensures that $(\pi(1), \pi(i))$ is distributed just as sampling $(a,b)$ condition on $s^\prime_a \neq s^\prime_b$. Using \cref{eq:mix_distribution},
            \begin{align*}
                \Pr_{r}[\algD(H_{r+1}^{s}) = 1] &= \Pr_{r}[\algD(H_{r}^{s}) = 1] \Pr[A] + \Pr_{r}[\algD(H) = 1] \Pr[B]\\
                &= \Pr_{r}[\algD(H_{r}^{s}) = 1] (1 - \Pr[B]) + \Pr_{r}[\algD(H) = 1] \Pr[B]
            \end{align*}
            Substitute in $\Pr_{r}[\algD(H_{r}^s) = 1] \geq \Pr_{r}[\algD(H_{r+1}^s) = 1] + \eps/4t$ (from \cref{cor:hybrid_advantange}),
            \begin{align*}
                \Pr_{r}[\algD(H_{r+1}^s) = 1] &\geq (\Pr_{r}[\algD(H_{r+1}^s) = 1] + \eps/4t)(1 - \Pr[B]) \\
                &\quad + \Pr_{r}[\algD(H) = 1] \Pr[B]\\
                \implies (\Pr[B] - 1)\eps/4t  &\geq (\Pr[B])(\Pr_{r}[\algD(H) = 1] - \Pr_{r}[\algD(H_{r+1}^s) = 1])
            \end{align*}
            Since $0 \geq (\Pr[B] - 1)\eps/4t$, so $\Pr[\algD(H_{r+1}^s) = 1]  \geq \Pr[\algD(H) = 1]$.
            \manualqed
        \end{proof}

\iflncs
\else
\subsection*{Acknowledgements}
Both authors are supported by the National Research Foundation, Singapore, under its NRF Fellowship programme, award no. NRF-NRFF14-2022-0010. AI tools were used as typing assistants for grammar and basic editing. We thank the EUROCRYPT
2026 reviewers for their helpful suggestions.
\fi

\iflncs
\bibliographystyle{splncs04.bst}
\else
\bibliographystyle{alpha}
\fi
\bibliography{refs}

@article{BBTV24,
  title={Near-Optimal Time-Sparsity Trade-Offs for Solving Noisy Linear Equations},
  author={Bangachev, Kiril and Bresler, Guy and Tiegel, Stefan and Vaikuntanathan, Vinod},
  journal={arXiv preprint arXiv:2411.12512},
  year={2024}
}

@article{CSZ24,
  title={Algorithms for Sparse LPN and LSPN Against Low-noise},
  author={Chen, Xue and Shu, Wenxuan and Zhou, Zhaienhe},
  journal={arXiv preprint arXiv:2407.19215},
  year={2024}
}

@inproceedings{RRS17,
  title={Strongly refuting random CSPs below the spectral threshold},
  author={Raghavendra, Prasad and Rao, Satish and Schramm, Tselil},
  booktitle={Proceedings of the 49th Annual ACM SIGACT Symposium on Theory of Computing},
  pages={121--131},
  year={2017}
}

@article{BEGKMZ22,
  title={Hidden progress in deep learning: Sgd learns parities near the computational limit},
  author={Barak, Boaz and Edelman, Benjamin and Goel, Surbhi and Kakade, Sham and Malach, Eran and Zhang, Cyril},
  journal={Advances in Neural Information Processing Systems},
  volume={35},
  pages={21750--21764},
  year={2022}
}

@inproceedings{GHKM23,
  title={Efficient algorithms for semirandom planted csps at the refutation threshold},
  author={Guruswami, Venkatesan and Hsieh, Jun-Ting and Kothari, Pravesh K and Manohar, Peter},
  booktitle={2023 IEEE 64th Annual Symposium on Foundations of Computer Science (FOCS)},
  pages={307--327},
  year={2023},
  organization={IEEE}
}

@inproceedings{Ale03,
  title={More on average case vs approximation complexity},
  author={Alekhnovich, Michael},
  booktitle={44th Annual IEEE Symposium on Foundations of Computer Science, 2003. Proceedings.},
  pages={298--307},
  year={2003},
  organization={IEEE}
}

@inproceedings{FGKP06,
  title={New results for learning noisy parities and halfspaces},
  author={Feldman, Vitaly and Gopalan, Parikshit and Khot, Subhash and Ponnuswami, Ashok Kumar},
  booktitle={2006 47th Annual IEEE Symposium on Foundations of Computer Science (FOCS'06)},
  pages={563--574},
  year={2006},
  organization={IEEE}
}

@inproceedings{ABW10,
  title={Public-key cryptography from different assumptions},
  author={Applebaum, Benny and Barak, Boaz and Wigderson, Avi},
  booktitle={Proceedings of the forty-second ACM symposium on Theory of computing},
  pages={171--180},
  year={2010}
}

@article{CHKV24,
  title={Somewhat Homomorphic Encryption from Linear Homomorphism and Sparse LPN},
  author={Corrigan-Gibbs, Henry and Henzinger, Alexandra and Kalai, Yael and Vaikuntanathan, Vinod},
  journal={Cryptology ePrint Archive},
  year={2024}
}

@article{App16,
  title={Cryptographic hardness of random local functions: Survey},
  author={Applebaum, Benny},
  journal={Computational complexity},
  volume={25},
  pages={667--722},
  year={2016},
  publisher={Springer}
}

@inproceedings{App12,
  title={Pseudorandom generators with long stretch and low locality from random local one-way functions},
  author={Applebaum, Benny},
  booktitle={Proceedings of the forty-fourth annual ACM symposium on Theory of computing},
  pages={805--816},
  year={2012}
}

@inproceedings{BSV19,
  title={XOR codes and sparse learning parity with noise},
  author={Bogdanov, Andrej and Sabin, Manuel and Vasudevan, Prashant Nalini},
  booktitle={Proceedings of the Thirtieth Annual ACM-SIAM Symposium on Discrete Algorithms},
  pages={986--1004},
  year={2019},
  organization={SIAM}
}

@inproceedings{AL16,
  title={Algebraic attacks against random local functions and their countermeasures},
  author={Applebaum, Benny and Lovett, Shachar},
  booktitle={Proceedings of the forty-eighth annual ACM symposium on Theory of Computing},
  pages={1087--1100},
  year={2016}
}

@article{Gol11,
  title={Candidate one-way functions based on expander graphs},
  author={Goldreich, Oded},
  journal={Studies in Complexity and Cryptography. Miscellanea on the Interplay between Randomness and Computation: In Collaboration with Lidor Avigad, Mihir Bellare, Zvika Brakerski, Shafi Goldwasser, Shai Halevi, Tali Kaufman, Leonid Levin, Noam Nisan, Dana Ron, Madhu Sudan, Luca Trevisan, Salil Vadhan, Avi Wigderson, David Zuckerman},
  pages={76--87},
  year={2011},
  publisher={Springer}
}

@article{AIK08,
  title={On pseudorandom generators with linear stretch in NC 0},
  author={Applebaum, Benny and Ishai, Yuval and Kushilevitz, Eyal},
  journal={Computational Complexity},
  volume={17},
  pages={38--69},
  year={2008},
  publisher={Springer}
}

@book{odonnell-book,
  title={Analysis of boolean functions},
  author={O'Donnell, Ryan},
  year={2014},
  publisher={Cambridge University Press}
}

@book{thomas-book,
author = {Cover, Thomas M. and Thomas, Joy A.},
title = {Elements of Information Theory (Wiley Series in Telecommunications and Signal Processing)},
year = {2006},
publisher = {Wiley-Interscience},
}

@inproceedings{Una23,
  author       = {Akin {\"{U}}nal},
  title        = {Worst-Case Subexponential Attacks on PRGs of Constant Degree or Constant
                  Locality},
  booktitle    = {Advances in Cryptology - {EUROCRYPT} 2023 - 42nd Annual International
                  Conference on the Theory and Applications of Cryptographic Techniques,
                  Lyon, France, April 23-27, 2023, Proceedings, Part {I}},
  volume       = {14004},
  pages        = {25--54},
  publisher    = {Springer},
  year         = {2023},
}

@InProceedings{ABCM25,
  author =	{Applebaum, Benny and Bui, Dung and Couteau, Geoffroy and Melissaris, Nikolas},
  title =	{{Structured-Seed Local Pseudorandom Generators and Their Applications}},
  booktitle =	{Approximation, Randomization, and Combinatorial Optimization. Algorithms and Techniques (APPROX/RANDOM 2025)},
  pages =	{63:1--63:26},
  year =	{2025},
  volume =	{353},
  publisher =	{Schloss Dagstuhl -- Leibniz-Zentrum f{\"u}r Informatik},
}

@inproceedings{BRT25,
author = {Bogdanov, Andrej and Rosen, Alon and Tan, Kel Zin},
title = {Sample Efficient Search to Decision for kLIN},
year = {2025},
publisher = {Springer-Verlag},
booktitle = {Advances in Cryptology – CRYPTO 2025: 45th Annual International Cryptology Conference, Santa Barbara, CA, USA, August 17–21, 2025, Proceedings, Part I},
pages = {203–220},
numpages = {18},
}

@inproceedings{CM01,
  title = {On {{Pseudorandom Generators}} in {{NC0}}},
  booktitle = {Mathematical {{Foundations}} of {{Computer Science}} 2001},
  author = {Cryan, Mary and Miltersen, Peter Bro},
  year = {2001},
  pages = {272--284},
  publisher = {Springer},
}

@inproceedings{MST03,
  title = {On E-{{Biased Generators}} in {{NC0}}},
  booktitle = {Proceedings of the 44th {{Annual IEEE Symposium}} on {{Foundations}} of {{Computer Science}}},
  author = {Mossel, Elchanan and Shpilka, Amir and Trevisan, Luca},
  year = {2003},
  pages = {136},
  publisher = {IEEE Computer Society},
}

@inproceedings{ABR12,
  title = {A Dichotomy for Local Small-Bias Generators},
  booktitle = {Proceedings of the 9th International Conference on {{Theory}} of {{Cryptography}}},
  author = {Applebaum, Benny and Bogdanov, Andrej and Rosen, Alon},
  year = {2012},
  pages = {600--617},
  publisher = {Springer-Verlag},
}

@article{AIK06,
  title = {Cryptography in \${{NC}}{\textasciicircum}0\$},
  author = {Applebaum, Benny and Ishai, Yuval and Kushilevitz, Eyal},
  year = {2006},
  journal = {SIAM Journal on Computing},
  volume = {36},
  number = {4},
  pages = {845--888},
}

@inproceedings{AM13,
  title = {Locally {{Computable UOWHF}} with {{Linear Shrinkage}}},
  booktitle = {Advances in {{Cryptology}} -- {{EUROCRYPT}} 2013},
  author = {Applebaum, Benny and Moses, Yoni},
  year = {2013},
  pages = {486--502},
  publisher = {Springer},
}

@inproceedings{BR11,
  title = {Input Locality and Hardness Amplification},
  booktitle = {Proceedings of the 8th Conference on {{Theory}} of Cryptography},
  author = {Bogdanov, Andrej and Rosen, Alon},
  year = {2011},
  pages = {1--18},
  publisher = {Springer-Verlag},
}

@inproceedings{BQ09,
  title = {On the {{Security}} of {{Goldreich}}'s {{One-Way Function}}},
  booktitle = {Proceedings of the 12th {{International Workshop}} and 13th {{International Workshop}} on {{Approximation}}, {{Randomization}}, and {{Combinatorial Optimization}}. {{Algorithms}} and {{Techniques}}},
  author = {Bogdanov, Andrej and Qiao, Youming},
  year = {2009},
  pages = {392--405},
  publisher = {Springer-Verlag},
}

@inproceedings{CEMT09,
  title = {Goldreich's {{One-Way Function Candidate}} and {{Myopic Backtracking Algorithms}}},
  booktitle = {Proceedings of the 6th {{Theory}} of {{Cryptography Conference}} on {{Theory}} of {{Cryptography}}},
  author = {Cook, James and Etesami, Omid and Miller, Rachel and Trevisan, Luca},
  year = {2009},
  pages = {521--538},
  publisher = {Springer-Verlag},
}

@inproceedings{CDMRR18,
  title = {On the {{Concrete Security}} of {{Goldreich}}'s {{Pseudorandom Generator}}},
  booktitle = {Advances in {{Cryptology}} -- {{ASIACRYPT}} 2018: 24th {{International Conference}} on the {{Theory}} and {{Application}} of {{Cryptology}} and {{Information Security}}, {{Brisbane}}, {{QLD}}, {{Australia}}, {{December}} 2--6, 2018, {{Proceedings}}, {{Part II}}},
  author = {Couteau, Geoffroy and Dupin, Aur{\'e}lien and M{\'e}aux, Pierrick and Rossi, M{\'e}lissa and Rotella, Yann},
  year = {2018},
  pages = {96--124},
  publisher = {Springer-Verlag},
}

@article{DMR23,
  title = {On the Algebraic Immunity---Resiliency Trade-off, Implications for {{Goldreich}}'s Pseudorandom Generator},
  author = {Dupin, Aur{\'e}lien and M{\'e}aux, Pierrick and Rossi, M{\'e}lissa},
  year = {2023},
  journal = {Des. Codes Cryptography},
  volume = {91},
  number = {9},
  pages = {3035--3079},
}

@inproceedings{LV17,
  title = {Limits on the {{Locality}} of {{Pseudorandom Generators}} and {{Applications}} to {{Indistinguishability Obfuscation}}},
  booktitle = {Theory of {{Cryptography}}: 15th {{International Conference}}, {{TCC}} 2017, {{Baltimore}}, {{MD}}, {{USA}}, {{November}} 12-15, 2017, {{Proceedings}}, {{Part I}}},
  author = {Lombardi, Alex and Vaikuntanathan, Vinod},
  year = {2017},
  pages = {119--137},
  publisher = {Springer-Verlag},
}

@inproceedings{OW14,
  title = {Goldreich's {{PRG}}: {{Evidence}} for {{Near-Optimal Polynomial Stretch}}},
  shorttitle = {Goldreich's {{PRG}}},
  booktitle = {Proceedings of the 2014 {{IEEE}} 29th {{Conference}} on {{Computational Complexity}}},
  author = {O'Donnell, Ryan and Witmer, David},
  year = {2014},
  pages = {1--12},
  publisher = {IEEE Computer Society},
}

@article{YGJL22,
  title = {Revisiting the {{Concrete Security}} of {{Goldreich}}'s {{Pseudorandom Generator}}},
  author = {Yang, Jing and Guo, Qian and Johansson, Thomas and Lentmaier, Michael},
  year = {2022},
  journal = {IEEE Trans. Inf. Theor.},
  volume = {68},
  number = {2},
  pages = {1329--1354},
}

@inproceedings{DV21,
  title = {From Local Pseudorandom Generators to Hardness of Learning},
  author = {Amit Daniely and Gal Vardi},
  year = {2021},
  pages = {1358-1394},
  booktitle = {Conference on Learning Theory, COLT 2021, 15-19 August 2021, Boulder, Colorado, USA},
  volume = {134},
  publisher = {PMLR},
}

@inproceedings{JLS22,
  title = {Indistinguishability {{Obfuscation}} from {{LPN}} over {{Fp}}, {{DLIN}}, and {{PRGs}} in {{NC0}}},
  booktitle = {Advances in {{Cryptology}} -- {{EUROCRYPT}} 2022: 41st {{Annual International Conference}} on the {{Theory}} and {{Applications}} of {{Cryptographic Techniques}}, {{Trondheim}}, {{Norway}}, {{May}} 30 -- {{June}} 3, 2022, {{Proceedings}}, {{Part I}}},
  author = {Jain, Aayush and Lin, Huijia and Sahai, Amit},
  year = {2022},
  pages = {670--699},
  publisher = {Springer-Verlag},
}

@inproceedings{JLS21,
  title = {Indistinguishability Obfuscation from Well-Founded Assumptions},
  booktitle = {Proceedings of the 53rd {{Annual ACM SIGACT Symposium}} on {{Theory}} of {{Computing}}},
  author = {Jain, Aayush and Lin, Huijia and Sahai, Amit},
  year = {2021},
  pages = {60--73},
  publisher = {Association for Computing Machinery},
}

@inproceedings{BCGIO17,
  title = {Homomorphic {{Secret Sharing}}: {{Optimizations}} and {{Applications}}},
  shorttitle = {Homomorphic {{Secret Sharing}}},
  booktitle = {Proceedings of the 2017 {{ACM SIGSAC Conference}} on {{Computer}} and {{Communications Security}}},
  author = {Boyle, Elette and Couteau, Geoffroy and Gilboa, Niv and Ishai, Yuval and Orr{\`u}, Michele},
  year = {2017},
  pages = {2105--2122},
  publisher = {Association for Computing Machinery},
}

@inproceedings{BCM23,
  title = {Sublinear-{{Communication Secure Multiparty Computation Does Not Require FHE}}},
  booktitle = {Advances in {{Cryptology}} -- {{EUROCRYPT}} 2023: 42nd {{Annual International Conference}} on the {{Theory}} and {{Applications}} of {{Cryptographic Techniques}}, {{Lyon}}, {{France}}, {{April}} 23--27, 2023, {{Proceedings}}, {{Part II}}},
  author = {Boyle, Elette and Couteau, Geoffroy and Meyer, Pierre},
  year = {2023},
  pages = {159--189},
  publisher = {Springer-Verlag},
}

@inproceedings{BCMPAR24,
  title = {Fast {{Public-Key Silent OT}} and~{{More}} from~{{Constrained Naor-Reingold}}},
  booktitle = {Advances in {{Cryptology}} -- {{EUROCRYPT}} 2024: 43rd {{Annual International Conference}} on the {{Theory}} and {{Applications}} of {{Cryptographic Techniques}}, {{Zurich}}, {{Switzerland}}, {{May}} 26--30, 2024, {{Proceedings}}, {{Part VI}}},
  author = {Bui, Dung and Couteau, Geoffroy and Meyer, Pierre and Passel{\`e}gue, Alain and Riahinia, Mahshid},
  year = {2024},
  pages = {88--118},
  publisher = {Springer-Verlag},
}

@inproceedings{ADINZ17,
  title={Secure Arithmetic Computation with Constant Computational Overhead},
  booktitle={CRYPTO},
  publisher={Springer},
  pages={223-254},
  author={Benny Applebaum and Ivan Damgård and Yuval Ishai and Michael Nielsen and Lior Zichron},
  year=2017
}

@inproceedings{RVV24,
  title = {Indistinguishability {{Obfuscation}} from~{{Bilinear Maps}} and~{{LPN Variants}}},
  booktitle = {Theory of {{Cryptography}}: 22nd {{International Conference}}, {{TCC}} 2024, {{Milan}}, {{Italy}}, {{December}} 2--6, 2024, {{Proceedings}}, {{Part IV}}},
  author = {Ragavan, Seyoon and Vafa, Neekon and Vaikuntanathan, Vinod},
  year = {2024},
  pages = {3--36},
  publisher = {Springer-Verlag},
}

@inproceedings{Fei02,
  title = {Relations between Average Case Complexity and Approximation Complexity},
  booktitle = {Proceedings of the Thiry-Fourth Annual {{ACM}} Symposium on {{Theory}} of Computing},
  author = {Feige, Uriel},
  year = {2002},
  pages = {534--543},
  publisher = {Association for Computing Machinery},
}

@inproceedings{DJ24,
  title = {Lossy {{Cryptography}} from~{{Code-Based Assumptions}}},
  booktitle = {Advances in {{Cryptology}} -- {{CRYPTO}} 2024: 44th {{Annual International Cryptology Conference}}, {{Santa Barbara}}, {{CA}}, {{USA}}, {{August}} 18--22, 2024, {{Proceedings}}, {{Part III}}},
  author = {Dao, Quang and Jain, Aayush},
  year = {2024},
  pages = {34--75},
  publisher = {Springer-Verlag},
}

@article{BCM25,
  author       = {Lennart Braun and
                  Geoffroy Couteau and
                  Kelsey Melissaris and
                  Mahshid Riahinia and
                  Elahe Sadeghi},
  title        = {Fast Pseudorandom Correlation Functions from Sparse {LPN}},
  journal      = {{IACR} Cryptol. ePrint Arch.},
  pages        = {1644},
  year         = {2025},
}

@inproceedings{COST19,
  title = {Expander-{{Based Cryptography Meets Natural Proofs}}},
  booktitle = {10th {{Innovations}} in {{Theoretical Computer Science Conference}} ({{ITCS}} 2019)},
  author = {Carboni Oliveira, Igor and Santhanam, Rahul and Tell, Roei},
  year = {2019},
  volume = {124},
  pages = {18:1--18:14},
  publisher = {Schloss Dagstuhl -- Leibniz-Zentrum f{\"u}r Informatik},
}

@inproceedings{AK19,
  title = {Sampling {{Graphs}} without {{Forbidden Subgraphs}} and {{Unbalanced Expanders}} with {{Negligible Error}}},
  booktitle = {2019 {{IEEE}} 60th {{Annual Symposium}} on {{Foundations}} of {{Computer Science}} ({{FOCS}})},
  author = {Applebaum, Benny and Kachlon, Eliran},
  year = {2019},
  pages = {171--179},
}

@mastersthesis{Zic17,
  author       = {Zichron, Lior},
  title        = {Locally computable arithmetic pseudorandom generators},
  school       = {School of Electrical Engineering, Tel Aviv University},
  year         = {2017},
  type         = {Master's thesis}
}

@inproceedings{Its10,
author = {Itsykson, Dmitry},
title = {Lower bound on average-case complexity of inversion of goldreich’s function by drunken backtracking algorithms},
year = {2010},
publisher = {Springer-Verlag},
booktitle = {Proceedings of the 5th International Conference on Computer Science: Theory and Applications},
pages = {204–215},
}

@article{AHI05,
author = {Alekhnovich, Michael and Hirsch, Edward A. and Itsykson, Dmitry},
title = {Exponential Lower Bounds for the Running Time of DPLL Algorithms on Satisfiable Formulas},
year = {2005},
publisher = {Springer-Verlag},
volume = {35},
number = {1–3},
journal = {J. Autom. Reason.},
pages = {51–72},
}

@article{BGPS06,
  title = {Randomized Gossip Algorithms},
  author = {Boyd, S. and Ghosh, A. and Prabhakar, B. and Shah, D.},
  year = {2006},
  journal = {IEEE Transactions on Information Theory},
  volume = {52},
  number = {6},
  pages = {2508--2530},
}

\end{document}